\documentclass{aa}  
\usepackage{graphicx}
\usepackage{txfonts}
\usepackage{multirow}
\usepackage[urlcolor=cyan, colorlinks=true, citecolor=blue, linkcolor=blue]{hyperref}

\begin{document} 
\title{Impact of stochastic star-formation histories and dust information on selecting quiescent galaxies with JWST photometry}
\titlerunning{Impact of stochastic SFHs \& dust on selecting QGs with JWST}
\authorrunning{K. Lisiecki}

    \author{
    K.~Lisiecki\inst{\ref{ncbj}}\and
    D.~Donevski\inst{\ref{ncbj},\ref{sissa}}\and
    A.~W.~S.~Man\inst{\ref{AMan}}, I.~Damjanov\inst{\ref{ID1}, \ref{ID2}}\and
    M.~Romano\inst{\ref{MR1}, \ref{MR2}}\and
    S.~Belli\inst{\ref{SB}}\and
    A.~Long\inst{9}\and
    G.~Lorenzon\inst{\ref{ncbj}}\and
    K.~Ma\l{}ek\inst{\ref{ncbj}}\and
    Junais\inst{\ref{J1}, \ref{J2}}\and
    C.~C.~Lovell\inst{\ref{CL1}, \ref{CL2}}\and
    A.~Nanni\inst{\ref{ncbj}}\and
    C.~Bertemes\inst{\ref{CB}}\and
    W.~J.~Pearson\inst{\ref{ncbj}}\and
    O.~Ryzhov\inst{\ref{OR}}\and
    M.~Koprowski\inst{\ref{MK}}\and
    A.~Pollo\inst{\ref{ncbj},\ref{UJ}}\and
    S.~Dey\inst{\ref{ncbj}}\and
    H.~Thuruthipilly\inst{\ref{ncbj}}
    }

\institute{
National Centre for Nuclear Research, Pasteura 7, 093, Warsaw, Poland; \email{krzysztof.lisiecki@ncbj.gov.pl}\label{ncbj} \and
SISSA, Via Bonomea 265, 34136 Trieste, Italy\label{sissa} \and 
Department of Physics \& Astronomy, University of British Columbia, 6224 Agricultural Road, Vancouver, BC V6T 1Z1, Canada \label{AMan} \and
Department of Astronomy and Physics, Saint Mary's University, 923 Robie Street, Halifax, NS B3H 3C3, Canada\label{ID1}\and
Canada Research Chair in Astronomy and Astrophysics (Tier II) \label{ID2} \and \
Max-Planck-Institut für Radioastronomie, Auf dem Hügel 69,
53121, Bonn, Germany \label{MR1}\and
INAF, OAPD, Vicolo dell’Osservatorio, 5, 35122 Padova, Italy \label{MR2} \and
Dipartimento di Fisica e Astronomia, Università di Bologna, via Gobetti 93/2, 40129 Bologna, Italy\label{SB}\and
Department of Astronomy, The University of Washington, Seattle, WA USA\label{AL}\and
Instituto de Astrof\'{i}sica de Canarias, V\'{i}a L\'{a}ctea S/N, E-38205 La Laguna, Spain\label{J1}\and
Departamento de Astrof\'{i}sica, Universidad de La Laguna, E-38206 La Laguna, Spain\label{J2}\and
Kavli Institute for Cosmology Cambridge, Madingley Road, Cambridge, CB3 0HA, UK\label{CL1}\and
Institute of Astronomy, University of Cambridge, Madingley Road, Cambridge CB3 0HA, UK\label{CL2}\and
Zentrum für Astronomie der Universität Heidelberg, Astronomisches Rechen-Institut Mönchhofstr, 12-14 69120 Heidelberg, Germany\label{CB}\and
Astronomical Observatory Institute, Faculty of Physics, Adam Mickiewicz University, ul. S\l{}oneczna 36, 60-286 Poznań, Poland\label{OR}\and
Institute of Astronomy, Faculty of Physics, Astronomy and Informatics, Nicolaus Copernicus University, Grudziądzka 5, 87-100 Toruń, Poland \label{MK}\and
Astronomical Observatory of the Faculty of Physics, Astronomy and Applied Computer Science, Jagiellonian University, ul. Orla 171, 30-244 Kraków, Poland \label{UJ}
}

   \date{Received 05.09.2025; Accepted 25.02.2026}

\abstract
{The James Webb Space Telescope (JWST) enables the identification of quiescent galaxies (QGs) out to early epochs, offering a transformative view of their evolution. However, photometric selection of quiescent galaxy candidates (QGCs) and the derivation of their key physical quantities, such as stellar masses (M$_\star$) and dust attenuation, remain highly sensitive to the assumed star-formation histories (SFHs), where dust–age degeneracies and modelling choices remain a major source of uncertainty.}
{
We aim to quantify how the inclusion of JWST/MIRI data and different SFH models impacts the selection and characterisation of QGCs. We test the robustness of the physical properties inferred from the spectral energy distribution (SED) fitting, such as M$_\star$, age, star formation rate (SFR), and dust attenuation (A$_V$), and study how they impact the quiescence criteria of the galaxies across cosmic time.}
{
We perform SED fitting for $\sim$13\,000 galaxies at $z\leq $\! 6 from the CEERS/MIRI fields with $\leq$20 optical-mid infrared (MIR) broadband coverage. We implement three SFH prescriptions: a flexible delayed, non-parametric, and an extended regulator model. For each SFH, we compare results obtained with and without MIRI photometry and dust emission models. We evaluate the impact of these configurations on both the number of QGCs, selected based on rest UVJ colours, sSFR and main-sequence offset, and on their key physical properties such as M$_\star$, A$_V$, and stellar ages.}
{The number of selected QGCs varies significantly with the choice of SFH ranging from 70 to 100, out of a mass-complete sample of $\sim 5\,000$ galaxies, depending on the model. This number increases to 103–180 when MIRI data are included, driven by improved constraints on both dust attenuation and M$_\star$. We find a strong correlation between A$_V$ and M$_\star$ of QGCs at $z\leq2.5$, with massive galaxies (M$_\star\sim10^{11}~\textrm{M}_\odot$) being $\sim\!1.5\!-\!4$ times more attenuated than low-mass galaxies (M$_\star\sim10^{9}~\textrm{M}_\odot$). Regardless of the SFH, $\sim13\%$ of QGCs exhibit significant attenuation (A$_V > 0.5$) in support of recent JWST results on dust-rich QGs.}
{}

   \keywords{Galaxies -- Galaxies: evolution -- Galaxies: quiescent -- Galaxies: dust}

   \maketitle
%

\section{Introduction}

Quiescent galaxies (QGs), often termed `evolved' or `passive', are characterised by minimal or absent star formation \citep{Strateva01, Willmer06, Franzetti07}, and lie below the star-forming main sequence \citep[MS; e.g.][]{Speagle14,Schreiber15,Wang22, Popesso23, Koprowski24, Merida25, Simmonds25}.
The physical mechanisms driving their quenching remain one of the most pressing puzzles in galaxy evolution \citep[][]{Faber07, Kriek08, Feldmann15,Man18, Zheng22, Akhshik23}, particularly with the discovery of massive QGs out to $z\sim5-7$ \citep[][]{Glazebrook17, Valentino20, Carnall23, Weibel2025}, which demands rapid quenching processes in the first $\sim$Gyr of cosmic history.
Key questions remain:
What is the interplay between quenching and the interstellar medium (ISM), especially dust?  To what extent can broadband photometry alone be used to constrain quenching timescales and the subsequent evolutionary pathways of QGs?

The advent of the James Webb Space Telescope (JWST) allows identifying high-$z$ QGs up to $z\sim7$ \citep[e.g.][]{Valentino23, Looser24, Weibel2025}, as well as QGs of lower stellar masses \citep[M$_\star<10^9$M$_\odot$;][]{Popesso23, Merida25}. The near-infrared (NIR) and mid-infrared (MIR) coverage of the JWST now enables the investigation of the age and dust-related parameters, which have long been difficult to disentangle due to the well-known degeneracy of dust-age-metallicity \citep[][]{Conroy13, Santini15}.
Inspecting this link has become an urgent topic, as recent studies reveal QGs with highly attenuated profiles \citep[e.g.][]{Setton24, Lu25, Bevacqua25} and/or a dust-enriched ISM \citep[e.g.][]{Gobat18, Whitaker21, Bezanson22, Morishita22, Donevski23, Lee24, Lorenzon25b} across a wide range of cosmic epochs ($0.4 \lesssim z \lesssim 4$), even though such systems are expected to be dust-poor at the time of observation.

Such discoveries force us to revisit our understanding of QG selection and the derivation of physical parameters from optical and near-IR photometry alone.
Identification of QGs has been done in different ways throughout the last decades, but mainly focussing on rest-frame colour-colour diagrams \citep[e.g.][]{Arnouts13}.
Such diagnostics allow us to break the degeneracy between dust reddening and stellar population ageing. 
They are especially powerful in selecting QG candidates (QGCs) from wide-field surveys, for which few photometric bands might be available.
The colour-colour diagrams are built with three rest-frame bands: one in near-ultraviolet (NUV) to optical, the second in optical range, and the final in NIR, in order to separate the population of QGs from dusty star-forming galaxies (SFGs).
Among many combinations, the $U-V$ versus $V-J$ (hereafter UVJ) diagram has been extensively tested for selection of QGs across all redshifts \citep[][]{Wuyts07, Williams09, Krywult17, Fang18, Akins22}, due to its correlations with specific star formation rate \citep[sSFR;][]{Williams10, Patel11, Martis19}, dust attenuation \citep[][]{Price14, Nagaraj22, Gebek25} and stellar ages \citep[][]{Whitaker13, Belli19}.
However, UVJ selection is known to be less reliable at $z \gtrsim 3$ where J band is not well sampled without rest-frame MIR data, which motivated the development of new colour selection techniques designed for higher redshifts \citep[][]{Antwi-Danso23, Gould23, Long24, Baker25}. 

Other commonly used methods are centred on main-sequence (MS) offsets and spectral tracers.
Spectral identification is the only reliable way to confirm the quiescent nature of a QGC, as it confirms the absence of ongoing star formation.
In particular, the Balmer break is a powerful indicator of the stellar population age, as it strongly correlates with the fraction of old vs. young stars \citep[][]{Balogh99,Kauffmann03, Haines17}. 
More precisely, absorption lines as \ion{Ca}{II} and H$\delta$ characterise the old stellar population that completely breaks the age-dust degeneracy \citep[][]{Bruzual83, Siudek17}.
Emission lines such as [\ion{O}{II}] and H$\alpha$ are instead used as tracers for very recent star formation \citep[][]{Kennicutt92, Villa-Velez21}.
Thus, the spectra encode direct information about stellar populations within galaxies and star formation histories \citep[SFHs;][]{Mathis06, Iyer19b, Iglesias-Navarro24}.
However, depending on the sample size and observing facilities, spectroscopy often requires long observing time and careful planning. 
In contrast, colour-colour diagrams, which can be obtained by fitting the spectral energy distribution (SED), sketch the dependence between age, dust attenuation and sSFR, and even offer a way to study quenching pathways \citep[e.g. fast vs. slow quenchers,][]{Tacchella22b}.
They allow select statistical samples of QGCs and take less time to implement in large areas of the sky with a wide range of redshifts \citep[e.g.][]{Daddi05, Arnouts07, Williams09, Ilbert13, Wu18}.

For years, most of the SED fitting codes modelled SFH with relatively simple analytical functions that depend on only a few parameters \citep[e.g.][]{Walcher11, Boquien20}.
The probe of the parameter space of the so-called parametric SFH is quick but lacks the flexibility to model more diverse SFHs e.g. with high burstiness or with a few episodes of rejuvenated star formation.
Recent studies introduce a new attempt to model SFHs with stochastic changes \citep[e.g.][]{Ocvirk06,Finlator07,Leja19, Iyer24, Annunziatella25, Carvajal-Bohorquez25}.

The methods proposed in these works overcome the simplicity of parametrised SFH without incurring significant additional computational cost. 
Furthermore, recent simulation-based studies define stochastic and physically motivated SFH models \citep[e.g.][]{Iyer17, Tacchella20}.
The new SFH models finally reach the flexibility needed to catch all changes shaping the evolution of a galaxy, but they were not tested on QGs.
Thus, the discussion about constraining SFHs, especially the quenching related timescales, with photometry-only studies is already ongoing \citep[][]{Smith15, Aufort24}.

In this work, we leverage JWST deep-field photometry to identify QGCs up to $z\sim\!6$ and, using SED fitting, to characterise their SFHs and key physical properties, such as stellar masses, rest-frame colours and dust attenuations. To achieve this goal, we implement and compare the NonParametric \citep{Leja19} and Extended Regulator \citep{Tacchella20} SFH models within Code Investigating GALaxy Emission \citep[CIGALE;][]{Boquien20}, testing their impact on derived galaxy properties.
We combine JWST MIR data with stochastic SFH to alleviate age and dust degeneracies.
We systematically evaluate how SFH choices, selection criteria, and the inclusion of MIR observations affect the physical and statistical properties of QGCs.
Finally, we study the quenching related timescales and their relation to the physical properties of QGCs within a purely photometric framework.

The paper is organised as follows. 
In Section~2, we describe the data sets and sample selection used in this study. Section~3 follows with a brief discussion of the adopted SFHs and our implementation of two new \textsc{CIGALE} modules based on stochastic SFHs, as well as our SED-fitting procedure. In Section~4, we present and discuss the results: we first compare how different SFH models affect the statistics of the selected QGCs, and then examine how their rest-frame UVJ colours and quenching timescales relate to key physical properties such as $M_\star$ and $A_V$.
In Section~5, we summarise our findings and conclude the study.

Throughout the paper, we assume the $\Lambda$CDM cosmological model with $H_0 = 70\,\textrm{km s}^{-1} \,\rm{Mpc}^{-1}$, $\Omega_m = 0.3$, and $\Omega_\Lambda = 0.7$.
We assume the \cite{Chabrier03} initial mass function (IMF) and conversion from other IMFs was performed using multiplicative factors from \cite{Madau14}. 

\section{Data}
We use the data collected from the Cosmic Evolution Early Release Science Survey \citep[CEERS; proposal No. 1345, P.I. Finkelstein;][]{Finkelstein22}.
CEERS covers an area of $\simeq94.6$ arcmin$^2$, previously observed for the 3D-HST survey in the AEGIS field \citep[][]{Brammer12}.
The CEERS programme makes use of both JWST imaging instruments, the Near-InfraRed Camera \citep[NIRCam;][]{Beichman12} and the MIRI \citep[][]{Bouchet15, Rieke15}, {with total area covered with MIRI observations reaching $\sim$28.28 arcmin$^2$.}
The abundance of MIR coverage in both the area and the number of photometric bands makes CEERS the optimal testing ground for stellar and dust properties of QGCs in a statistical manner, and no other field provides a similar advantage in the study of QGCs evolution.

\subsection{HST--JWST/NIRCam photometry}
To build our parent catalogue, we use seven NIRCam filters sampled to a pixel size of 0.03 arcsec, out of which six broad-band (F115W, F150W, F200W, F277W, F356W and F444W) and one medium-band (F410W), and seven broad-band MIRI filters sampled to pixel size of 0.09 arcsec (F560W, F770W, F1000W, F1280W, F1500W, F1800W and F2100W).
We complement JWST data with six Hubble Space Telescope (HST) broad-band filters sampled to a pixel size of 0.03 arcsec (F606W, F814W, F105W, F125W, F140W and F160W; \citealt{stefanon17}). By combining JWST with HST data, we obtain the wealthy optical to mid-infrared coverage spanning 20 bands. We adopt the HST and JWST NIRCam fluxes from the ASTRODEEP-JWST catalogue \citep{Merlin24}, which complements NIRCam observations with archival HST observations.
The catalogue consists of 82\,547 sources, and its our parent sample.
We take advantage of the homogeneous reduction and flux extraction for both NIRCam and HST observations within the catalogue. 
Additionally, to keep our sample consistent, we use photometric redshifts calculated by \cite{Merlin24}\footnote{MIRI data that were not used in photo$-z$ estimation. We provide a test of the influence of MIRI data on photo-$z$ estimation in App.~\ref{APP:photoz}, showing that there is no systematic shift.}.

For a detailed description of flux extraction and analysis, we refer to \cite{Merlin24}.
In short, after reprojecting all the images (NIRCam and HST) to the same pixel grid, the F356W and F444W images were stacked and used as a detection image. 
SExtractor \citep[][]{Bertin96} was used for the detection and flux derivation. 
The errors were estimated through root mean square (RMS) maps.
Comparison with spectroscopic redshifts ($z_{spec}$) shows that only $\sim$ 6\% of the sample has a redshift mismatch greater than 0.15 \citep[see Fig. 12 in][]{Merlin24}.

The catalogue is 90\% complete at $\sim$29.04 mag in stacked F356W and F444W filters \citep[Fig.~\ref{fig:massCompletenessAll}; see also Fig.~4 in][]{Merlin24}. 
According to our analysis (App.~\ref{app:completeness}) it translates to 90\% completeness at M$_*\sim10^{7.9}$M$_\odot$ at $z=6$ for the entire catalogue.
The completeness for the QG-only sample reaches a similar depth, while for star-forming galaxies only, we estimate 90\% completeness at M$_*\sim10^{7.6}$M$_\odot$ for the same redshift.

\subsection{MIRI photometry}\label{SEC:MIRIPhotometry}
We perform optimised extractions of MIRI fluxes across all available MIRI bands.
The detailed description of MIRI flux estimates, their associated errors, and inferred upper limits, can be found in the App.~\ref{APP:MIRI}. 

Briefly, we extract total fluxes from MIRI sources in each of the pointings using SExtractor.
We use the MIRI mosaics version 0.6 data release \citep[][]{yang23}, which has corrected astrometry based on the HST reference.
We follow \cite{yang23} and use the MAG\_AUTO parameter for flux measurements. We cross-match the NIRCam/HST catalogue from \cite{Merlin24} with our MIRI fluxes within a 0.5-arcsec radius. 
In our catalogue, and further SED fitting, we put a 5$\sigma$ upper limit for MIRI fluxes when no direct detection (no source in crossmatch radius or signal to noise $<3$) is found.
We use 5$\sigma$ instead of 3$\sigma$ since our analysis of upper limits was performed for point-like sources, and galaxies are spatially resolved.
The upper limits values vary, depending on the band and the MIRI pointing, from the deepest 26.75 mag for F770W pointing 3, to 23.64 mag for F2100W pointing 1 (see the full list of 5$\sigma$ AB upper limits in Tab.~\ref{TAB:MIRI_upperlimits}).

\subsection{Final sample}\label{SEC:FinalSample}
We aim to understand how the inclusion of MIR points (crucial to assess age-dust degeneracy), combined with the SFH model choice, affects the selected QGCs and their physical properties.
For this reason, we restrict the parent CEERS catalogue to include only sources within the CEERS/MIRI footprint covering $\sim$28.28 arcmin$^2$.

This limitation results in a sample in which all our sources have been observed in at least one MIRI band. Since the MIRI footprints of different filters are not equal (see the coverage of pointings in Tab.~\ref{TAB:MIRI_upperlimits}), the number of filters covering a single source can vary from one to six. 
This choice restricts us to 12\,939 galaxies for which 1\,939 have direct MIRI detection above 3$\sigma$. The remaining 11\,000 have at least one MIRI upper limit.\footnote{We share our full photometric catalogue of HST/NIRCam/MIRI sources at our GitHub repository (github.com/lisieckik/CIGALE\_QGs).} We show the redshift distribution of these galaxies in Fig.~\ref{fig:massCompletenessAll}.

Finally, throughout this work, we focus only on a mass-complete sample.
Thus, we choose to study only galaxies with stellar masses above the 90\% completeness limit, M$_\star\geq 10^8$M$_\odot$, in all three SED fitting \textit{MIRI runs} (see Sec.~\ref{sec:SEDmodeling}). Our final sample consists of 5\,021 galaxies at $0.1\lesssim z\lesssim 6$ out of which 1\,574 have at least one detection in the MIRI bands\footnote{We note that  2\,614 objects reside at $z<2.5$, where MIRI fluxes probe the emission from warm dust and PAH.}.
We have: 572, 832, 41, 117, 3, and 9 sources observed in 1, 2, 3, 4, 5, and 6 MIRI bands in final sample, respectively.

\begin{figure*}[ht]
\centering
\includegraphics[width = 0.98\textwidth]{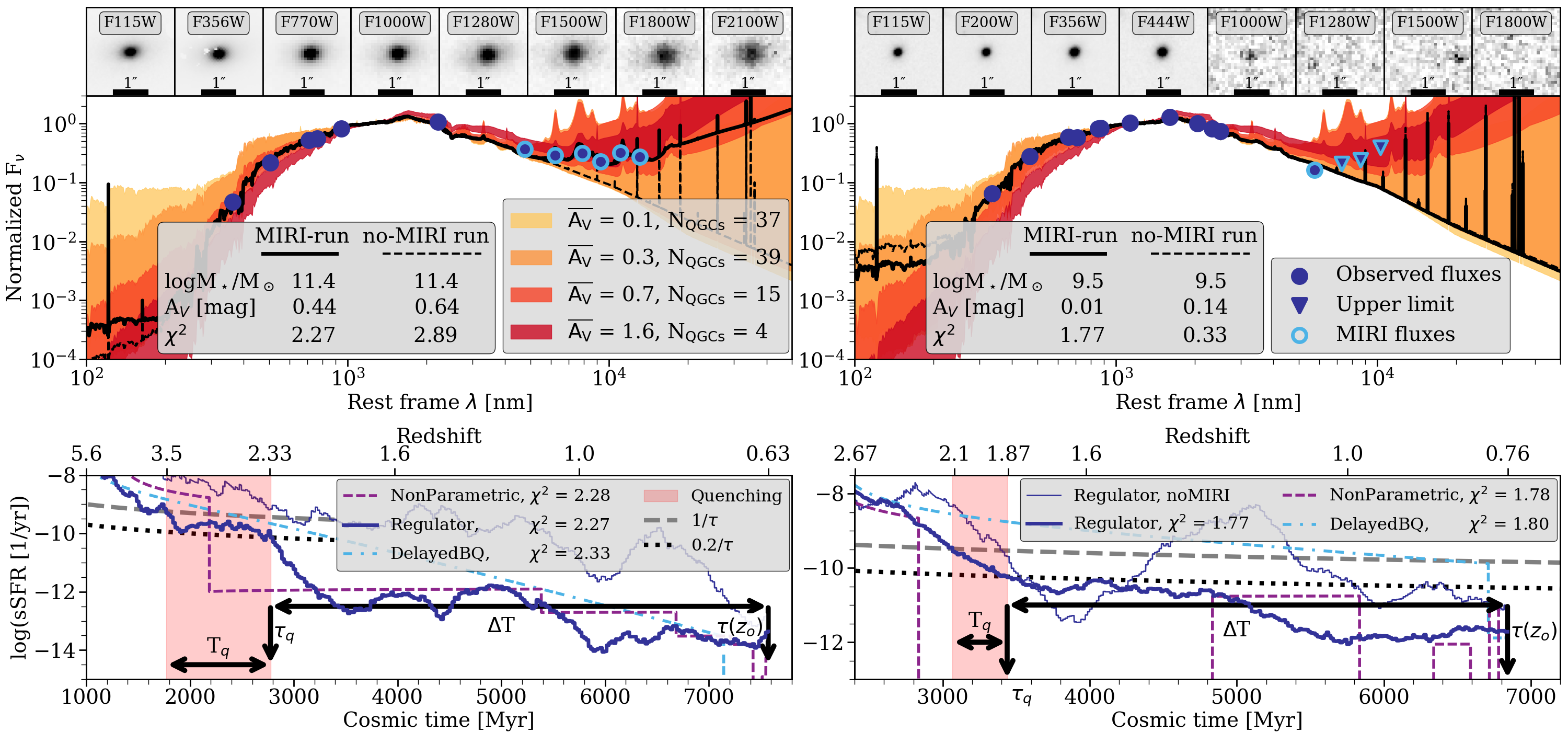}
\caption{
Exemplary QGCs from our final sample: a massive system on the left; a less–massive one on the right.
\textit{Top:} CIGALE SEDs. Black solid and dashed lines show the best–fit SEDs for the same galaxy using the Regulator SFH with and without JWST/MIRI photometry for 
\textit{MIRI} and \textit{no–MIRI run}, respectively. Blue circles are detections; triangles are $3\sigma$ upper limits. JWST cutouts are displayed above each SED. Shaded envelopes indicate the range of $K$–band–normalized SEDs for QGCs binned by median $A_V$; the legend lists the median $A_V$ values and number of objects in each bin.
\textit{Bottom:} Specific star-formation histories (sSFHs) for the same QGC from the \textit{MIRI} run. The dark–blue solid, violet dashed, and light–blue dash-dotted lines correspond to the \textit{Regulator}, \textit{NonParametric}, and \textit{DelayedBQ} models, respectively. The thick and thin dark-blue lines show the difference between \textit{MIRI} and \textit{no-MIRI runs}. The black dotted line marks the quiescent threshold and the black dashed line the star-forming threshold (following \citealt{Pacifici16}). Arrows mark the quantities used in this work for the \textit{Regulator} model: the quenching time $T_q$, the quenching moment $\tau_q$, the cosmic time at the observation redshift $\tau(z_o)$, and the time since quenching $\Delta T \equiv \tau(z_o)-\tau_q$.}
\label{fig:examplesSFH}
\end{figure*}

We are aware that, the AGN dust emission can contribute to MIRI photometry (e.g. \citealt{yang23a}.) Therefore, we conducted a test to evaluate its impact on our results.  We cross-match the final sample with the 42 AGN candidates identified in the CEERS/MIRI observations (pointings 1, 2, 5, and 8), as described in \cite{Chien24}, resulting in 24 direct counterparts. 
The AGN contamination in our QGC sample (see Sect. \ref{SEC:QGsSelectionEff}) is approximately 4\%. Furthermore, none of the AGN contaminants in our QGCs exceeds a V-band attenuation of 0.5 mag.
Nevertheless, we remove these known contaminant sources from any QGCs sample in the following analysis.

\section{Spectral energy distribution modelling}
We use CIGALE \citep[version 2022.01;][]{Boquien20}, as it is optimised for broad band photometry, to model and fit the SED of observed galaxies. 
CIGALE uses energy balance conservation between the dust-absorbed stellar emission and its re-emission in the infrared (IR). 
It goes over a grid of models, with parameter values set by the user, and identifies the best fit by minimising chi square ($\chi^2$).
We take advantage of the modularity of CIGALE and implement stochastic SFHs within the framework (see Sec.~\ref{SEC:nonparam} and Sec.~\ref{SEC:regulator}). 

\subsection{Star formation histories}
With respect to SFHs in CIGALE, there are two main approaches implemented. 
The first one is straightforward using previously prepared SFH, i.e. templates or directly from simulations, as input to test what SED will be produced with such SFH.
The second is built on parametric models.
Parametric SFHs assume that star formation in a galaxy is a continuous process, which can be described by a few parameters like age and the e-folding time of the main stellar population. 
For a detailed description of all available SFH parameterisations, we refer the reader to \cite{Boquien20}.

For the purpose of this study, we implemented the diverse stochastic SFHs within CIGALE, namely NonParametric \citep[][]{Leja19,Carnall19} and Extended Regulator \citep[Regulator hereafter;][]{Tacchella20}.
The description of the implementation, technical details, as well as the original method can be found in App.~\ref{APP:implementationSFH}. 
We also share the implementation in the form of a public GitHub repository\footnote{github.com/lisieckik/CIGALE\_QGs}.
Throughout the study, we provide detailed comparison of results achieved with three SFH models: parametric flexible Delayed + Burst/Quenching model (DelayedBQ hereafter), NonParametric and Regulator.
The visual comparison of specific SFH models (sSFH hereafter) prepared with new modules used in this work is shown in the bottom panel of Fig.~\ref{fig:examplesSFH}.

In the NonParametric SFH model, the SFR changes randomly between defined time steps. 
The changes follow the Student-t distribution.
We follow the implementation described by \cite{Leja19} with prior continuity.
This approach results in reduced burstiness, that is, fewer rapid changes in SFR in time.
In our implementation, we allow for five parameters (an exemplary set of parameters can be found in Tab.~\ref{TAB:CIGALE_inp}) in this module.

For the Regulator model, we follow the approach described in \cite{Tacchella20} and \cite{Iyer24}.
The Regulator SFH model is based on the control of the SFR by the assigned gas mass reservoir.
The stochastic nature of the SFR is a result of 1) gas inflow/outflow rates, 2) gas cycling in equilibrium between atomic and molecular states, and 3) the formation, lifetime and disruption of giant molecular clouds (GMCs). 
In our implementation, we allow for eight parameters in the Regulator model (an exemplary set of parameters can be found in Table~\ref{TAB:CIGALE_inp}).

DelayedBQ \citep{Ciesla16, Ciesla21} is a flexible extension of the classical delayed-$\tau$ model.
By adding an instantaneous quench/burst, the DelayedBQ model is flexible for better sensitivity to SFR within recently quenched galaxies.
The DelayedBQ model has the following form:
\begin{equation}
\textrm{SFR}(t) \propto \left\{
    \begin{aligned}
        & t\times\exp{(-t/\tau_{main})}, \mathrm{when}\,\, t\leq t_{flex} \\
        & r_{\textrm{SFR}}\times\textrm{SFR}(t=t_{flex}), \mathrm{when}\,\, t>t_{flex}
    \end{aligned}
    \right.,
\end{equation}
where $t$ is the time, $\tau_{main}$ is the e-folding time, $t_{flex}$ is the time at which star formation is affected by instantaneous change, and $R_{\textrm{SFR}}$ is the ratio between SFR($t>t_{flex}$) and SFR($t=t_{flex}$):
\begin{equation}
    r_{\textrm{SFR}} = \frac{\textrm{SFR} (t>t_{flex})}{\textrm{SFR}(t = t_{flex})}.
\end{equation}

\subsection{SED fitting runs and parameter space} \label{sec:SEDmodeling}
To study the influence of the SFH model on the estimated physical properties, such as M$_\star$, SFR or V-band attenuation (A$_V$), we perform three runs: 1) with the parametric DelayedBQ model, 2) with the NonParametric model and 3) with the Regulator model.
We keep all parameters not related to SFH the same between runs. 
The only ones that are different are the SFH variables that are module-dependent and the stellar population ages (depending on redshift).
We summarise in the following the main methods and parameters used for our models, with explicit values detailed in the App.~\ref{APP:Cigale}.

For the NonParametric module, we use \textit{nLevels} = 8, since it was tested that robust results can be achieved with $4\leq$ \textit{nLevels} $\leq 14$ \citep[][]{Leja19, Tacchella20, Lower20}.
The \textit{lastBin} is set to 30~Myr following \cite{Leja19} and \cite{Ciesla23}.
We make in total five iterations of each run with both stochastic SFH modules to test the stability of the results (see App.~\ref{APP:stabilityOfPhysics}).
In each iteration, we set \textit{nModels} to 200.
In total, we generate 1\,000 random sampled SFH models, each of them is tested in CIGALE.
Thus, we have not only the best model (found by the least reduced $\chi^2$), but also a group of five iterations of the best models per stochastic SFH. 
In other words, each galaxy has five independent solutions within the same stochastic SFH model framework.
We take advantage of that by checking the stability of the solution.
Whenever we use all five solutions, we mention it explicitly in the text.
According to our tests, splitting the CIGALE run into five smaller runs to sample more SFH models does not influence the results. More details can be found in App.~\ref{APP:stabilityOfPhysics}.

For the Regulator module, we set \textit{nLevels} at 300.
The number of time-steps in the Regulator model is not limited as much as in NonParametric, since the SFR changes are physically motivated.
Although this number is arbitrary, it results in a time resolution that is always better than $\sim$45 Myr per step (depending on the formation redshift and the observation redshift).
This is enough to distinguish, for example, fast quenchers (quenching time $<100$ Myr) from slow quenchers \citep[][]{Belli19, Tacchella22b}.
The \textit{nModels} parameter, similarly to the NonParametric run, is tested in five iterations by 200 each, in total 1\,000 SFH models.
The parameters regulating variability and time scales, \textit{sigmaReg, tauEq, tauFlow, sigmaDyn} and \textit{tauDyn}, were set at 1, 1300, 125, 0.07 and 35, respectively. 
These values are between the Milky Way analogue and the cosmic noon galaxy from \cite{Iyer24}.
Since we expect a diverse galaxy population in our sample, we decide to make it a compromise, as checking all combinations is out of our computational capabilities and out of the scope of this study.

We are aware of high burstiness reported in high-$z$ ($z>2.5$) and/or low-M$_\star$ ($\log$M$_\star/$M$_\odot<10^9$) galaxies \citep[e.g.][]{Cole25, Mintz25}.
To rule out its importance for our analysis, we conducted additional tests for a lower number of galaxies, with \textit{sigmaReg} = 1, 1.5, 2, 3.
We report that the results (M$_\star$, A$_V$, SFR and age) found with different values of \textit{sigmaReg} are consistent within the uncertainties reported by CIGALE. 
Thus, we expect this parameter to have a low-influence on the following analysis.

\begin{figure}[ht]
\centering
\includegraphics[width = 0.48\textwidth]{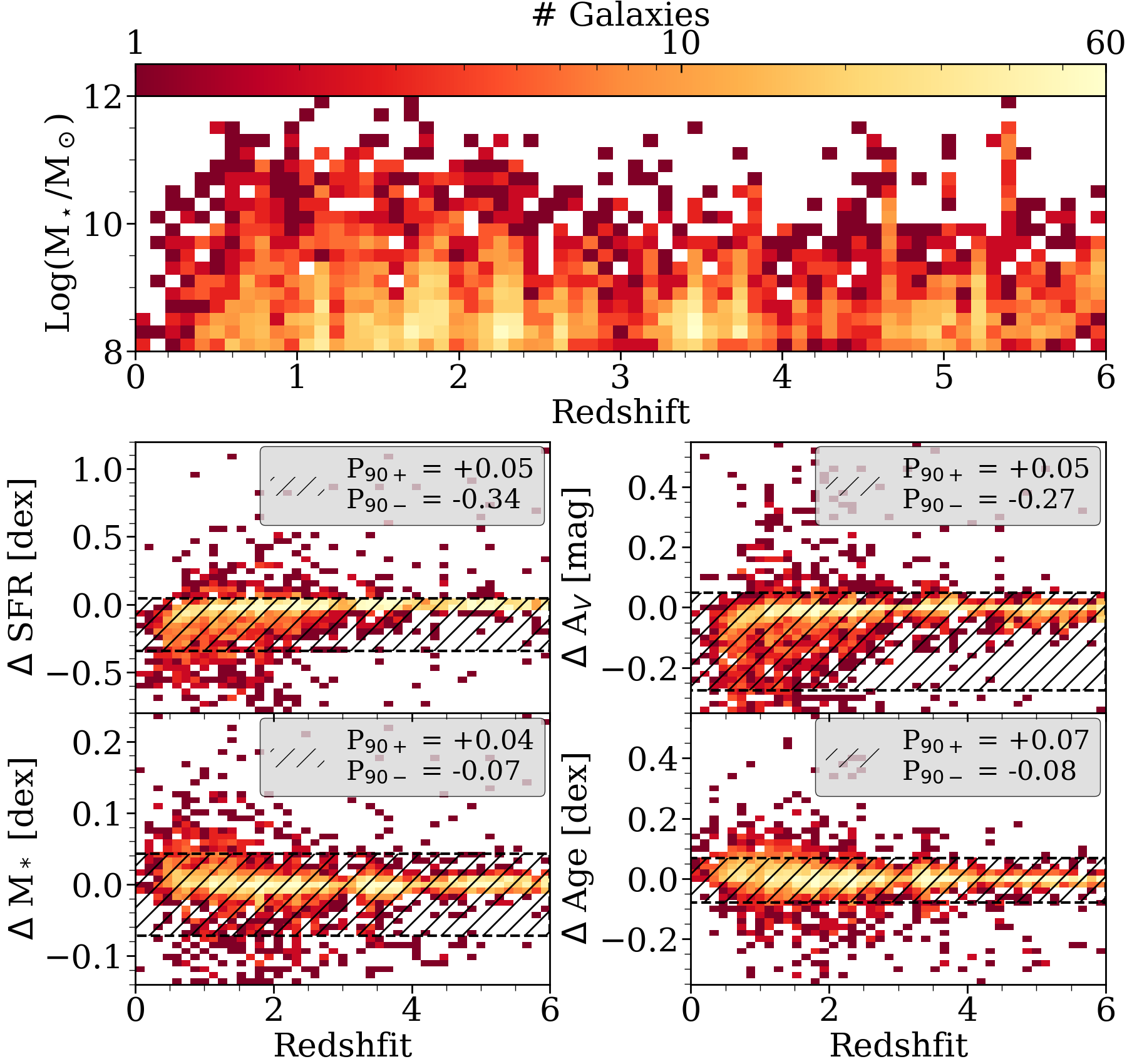}
\caption{
Top: the distribution of the M$_\star$ in function of redshift of the final sample from Regulator MIRI run.
Bottom: distributions of differences in main physical properties (SFR in M$_\odot/$yr and the M$_\star$ in M$_\odot$, A$_V$ in mag, age in Myr) between \textit{MIRI} and \textit{no-MIRI runs}. The hatched region shows the 90th percentiles of the distributions (P$_{90}$) in both directions.
}
\label{fig:ResultsHist}
\end{figure}

In all runs, we assume \cite{Bruzal03} single stellar population models with the IMF given by \cite{Chabrier03}.
We use the \cite{Charlot00} attenuation module, as our main interest is to study how the inclusion of MIRI data reflects the SFH output.
This module assumes two separate attenuation curves for young and old stellar populations.
Finally, we use the dust emission model described in \cite{Draine14}.

Since our sample covers a broad redshift range (${0<z<6}$), we split it into five sub-samples with lower redshift coverage.
The main purpose is to keep the computation relatively easy, with as little change in input parameters as possible.
The number of galaxies within each sub-sample, redshift ranges of the sub-samples, and their influence on the input parameters are shown in the first part of Tab.~\ref{TAB:CIGALE_inp}.

To test the influence of the dust component and MIR observations on the results, we perform a second set of CIGALE runs using the same final sample, but without observed MIRI data (fluxes and/or upper limits). 
Since here we restrict our data to NIR, we also disable the dust emission model to speed up the calculations. 
All other parameters stay the same. 
From now on, the main runs with inclusion of MIRI data and dust emission module will be referred to as \textit{MIRI runs}.
The second set, with no MIRI data and dust emission module included will be referred to as \textit{no-MIRI runs}. 
The exemplary SEDs for both \textit{MIRI} and \textit{no-MIRI runs} are presented in the upper panel of Fig.~\ref{fig:examplesSFH}.

Throughout the analysis, we utilise the Bayesian values of each property estimated by CIGALE (through an internal Bayesian procedure), unless explicitly written otherwise.
The M$\star$-$z$ distribution of the final sample for the Regulator SFH model is presented in the upper panel in Fig.~\ref{fig:ResultsHist}.
Furthermore, redshift and M$_\star$ histograms are presented in Fig.~\ref{fig:massCompletenessAll}.
The photo-z errors are not propagated.

\section{Results and discussion}\label{sec:ResultsDiscuss}

In this section, we investigate and discuss the influence of SFH, MIRI data and selection criteria on the number of QGCs selected. 
This is followed by a comprehensive analysis of capabilities to constrain the SFHs with the photometry-only SED fitting. 
Finally, we study the inferred quenching-related properties and dust attenuation evolution after quenching.

We focus our study on four fundamental global physical properties: M$_\star$, the star formation rate (SFR; averaged over 10 Myr), A$_V$, and the age (look-back time of formation age, it is the time since the onset of star formation).
To quantify the stability of the physical properties, we use the coefficient of variation (CV, relative standard deviation), $\textrm{CV} = \sigma/\mu$, where $\sigma$ is the standard deviation (calculated from five iterations) and $\mu$ is the value of the physical property such as M$_\star$, SFR or A$_V$ of the iteration with the lowest $\chi^2$. 

As a sanity check, we test the stability of the results with stochastic SFH within the final sample.
Since in each iteration all SFH models are different, the best result found by CIGALE will vary, contrary to that of the parametric SFH. 
The resulting scatter in the inferred physical properties, such as M$_\star$ or SFR, is quantified in App.~\ref{APP:stabilityOfPhysics}.
We report high consistency between iterations for both NonParametric and Regulator models for M$_\star$ (on average, M$_\star$ in each iteration differ by less than 6\%), A$_V$ (below 8\%) and age (below 6\%).
Similar consistency is found for SFR but only in the Regulator model (below 3\%).
The SFR in the NonParametric can vary up to $\sim30$\%. 
However, this variation is mainly due to low SFR values, which does not influence the overall selection.

\subsection{Influence of MIRI data on inferring physical properties}\label{Sec:MIRIInfluence}
We further test how incorporating MIRI data and dust emission models in the SED fitting affects the results, compared to SED fits that exclude both.
The comparisons of the distributions from \textit{no-MIRI} and \textit{MIRI runs}, for Regulator SFH used as reference model, are presented in Fig.~\ref{fig:ResultsHist}.

\begin{figure}[ht]
\centering
\includegraphics[width = 0.49\textwidth]{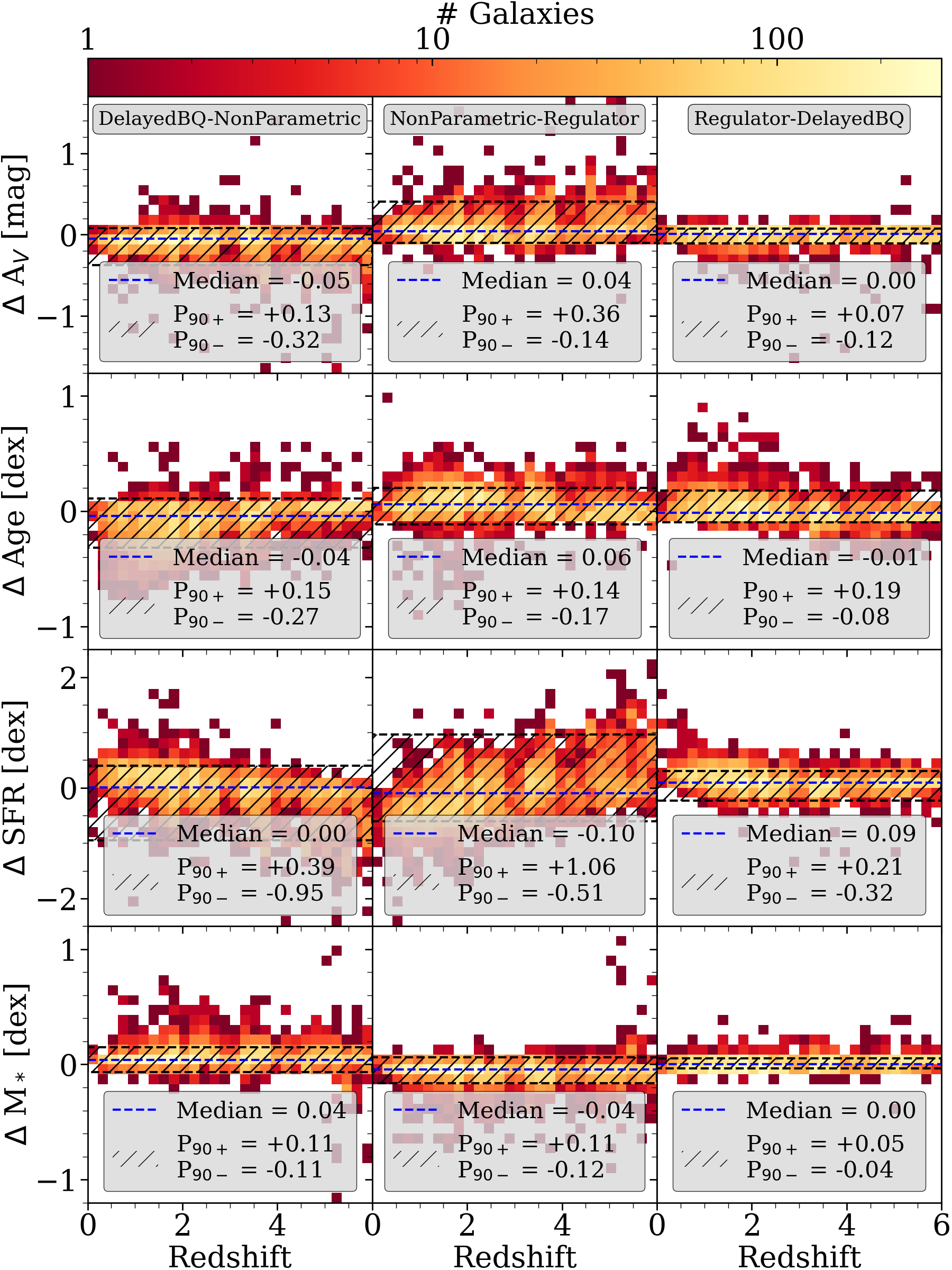}
\caption{Difference in estimated physical properties of galaxies from the final sample using different SFH models within \textit{MIRI run} as a function of redshift. Panels from left to right: DelayedBQ-NonParametric, NonParametric-Regulator and Regulator-DelayedBQ. Panels from top to down show distribution of difference for: A$_V$ in mag, age in Myr, SFR in M$_\odot/$yr and the M$_\star$ in M$_\odot$. The blue dashed line presents the median of distribution. The hatched region shows the 90th percentiles of the distributions (P$_{90}$) above and below the median value.}
\label{fig:Physical_diff}
\end{figure}

In general, we find that the inclusion of MIRI measurements has a negligible effect on the derived M$_{*}$ (P$_{90}<0.09$ dex) and the ages (P$_{90}<0.1$ dex), which remain consistent in all SFH.
This was confirmed for high-mass galaxies \citep[M$_\star>10^{10}$M$_\odot$;][]{Wang25}, but we find similar accuracy for lower-mass galaxies as well.
However, it plays a crucial role in the constraint of SFR and A$_V$. 
This SFR-A$_V$ degeneration is expected when no IR coverage is available \citep[see e.g.][]{Riccio21, Malek24}, since CIGALE uses the energy budget balance.
In our case, without MIRI, the SFRs are systematically overestimated up to 0.65 dex (with a median $\sim0.1$dex) due to poorer constraints on dust emission, which regulates energy balance. 
Thus, the QGC samples are more complete in \textit{MIRI runs}.
Similarly, in the absence of MIR coverage, A$_V$ exhibits a small but steady asymmetric bias towards higher values, up to 0.25 mag (with a median $\sim0.06$mag). 

These results highlight the importance of MIRI to better constrain the IR side of the SED (thus energy balance) and help overcome degeneracies between dust and age in QGs.
By including MIR points and dust emission, we put a soft prior on the energy absorbed by dust, thus SFR.
However, we must mention that while JWST clearly helps to overcome the degeneracy between dust and age, there is still the degeneracy between metallicity and age \citep[e.g.][]{Cheng25}.
This will be explored with spectroscopic data in future work.

It is worth noting that galaxies with at least one MIRI detection from our final sample are generally lower$-z$ with a median $z\sim1.8$, compared to $z\sim2.4$ for the entire final sample.  
Similarly, these galaxies have higher M$_\star$ (median at $10^{9.2}$ M$_\odot$) compared to the final sample ($10^{8.7}$ M$_\odot$).
This is caused by shallower observations with MIRI compared to NIRCam.
The distributions of the other parameters do not vary significantly.

\subsection{Impact of SFH modelling on derived physical properties}\label{sec:imapctSFH}

To understand the biases introduced by the choice of the SFH model in deriving M$_\star$, A$_V$, SFR and age of QGCs, we first study the influence of SFH on the whole sample. 
We utilize the \textit{MIRI run}.
In Fig.~\ref{fig:Physical_diff} we compare the physical quantities of the SED runs with different SFH models.
We report great agreement for all properties apart from SFR with NonParametric SFH model.
The estimation of M$_\star$ and age are consistent up to $\sim$0.15 dex and A$_V$ up to $\sim$0.3 mag between SFH models. 
The SFR estimated with DelayedBQ and Regulator models agree up to $\sim$ 0.4 dex.
The SFR stability between iterations with the use of the NonParametric model is low, especially for negligible SFR values (see App.~\ref{APP:stabilityOfPhysics}).
The strong underestimation of SFR by NonParametric model is prominent only in low-SFR galaxies and we can suspect the overall consistency up to $\sim$ 0.5 dex compared to other SFH models.
A similar underestimation of SFR, reaching $\sim0.3$dex, in low-SFR galaxies (logSFR$<0$) has been shown by comparing direct cosmological simulation results with the CIGALE SED fitting of mock observations of the same galaxies \citep[see Figure 4 from][]{Arango-Toro25}.
Additionally, \cite{Annunziatella25} showed that the NonParametric model favours a faster mass assembly than parametric models.
This may result in earlier quenching (compared to parametric SFHs models), followed by lower values of SFR. 
The comparison between Regulator and NonParametric SFH models correlates well with the results of \cite{Wan24}.

Across the three SFH methods, the dominant source of discrepancy arises from fundamentally different constraints on the accessible individual models. 
Parametric models such as DelayedBQ enforce smooth evolutionary shapes with a single sharp feature such as quenching or burst.
The NonParametric approach allows for arbitrary temporal structure and consequently reconstructs more complex or discontinuous histories when the data permit.
The Regulator SFH restricts solutions to be consistent with gas-cycling physics, tying the SFR evolution to inflow, outflow, and efficiency prescriptions. 
As a result, each method interprets the same observational features through a different optics.
These factors lead to systematic mismatches in inferred SFRs (see Fig.~\ref{fig:Physical_diff}) and mass-assembly timescales \citep[e.g.,][]{Annunziatella25}.

\subsection{Photometric selection of QGCs: impact of SFH models}\label{SEC:QGsSelectionEff}

In the next step, we test the effect of the selection criteria on the resulting QGCs sample for each SFH model in both \textit{MIRI} and \textit{no-MIRI} CIGALE runs.
We test various popular and empirically defined methods for selecting QG, such as: MS \citep[][]{Popesso23, Koprowski24, Merida25}, the criterion based on specific SFR (sSFR; $\textrm{sSFR}\equiv \textrm{SFR/M}_*$) proposed by \cite{Pacifici16} and UVJ selection with correction for $z>4$ sources \citep[][]{Antwi-Danso23}.
Using more than one definition of QGC will show us whether the SFH model is more influential than the QGC selection criterion.
We report the effect of the selection criteria on the resulting QGCs sample for each SFH model in both \textit{MIRI} and \textit{no-MIRI} CIGALE runs.
The comparison of the number of QGCs selected with different SFH and criteria is shown in Fig.~\ref{fig:nQGs}.
We find that both choice of SFH and inclusion of MIRI data strongly impact photometric selection of QGCs, but interestingly these effects appear to be independent of each other.
We normalise all selection criterion to the \cite{Chabrier03} IMF.
Our prime attention in the following analysis would be to the most conservative criterion, which is the one based on specific SFR (see below).

\subsubsection{Main sequence selection}

\begin{figure}[ht]
\centering
\includegraphics[width = 0.48\textwidth]{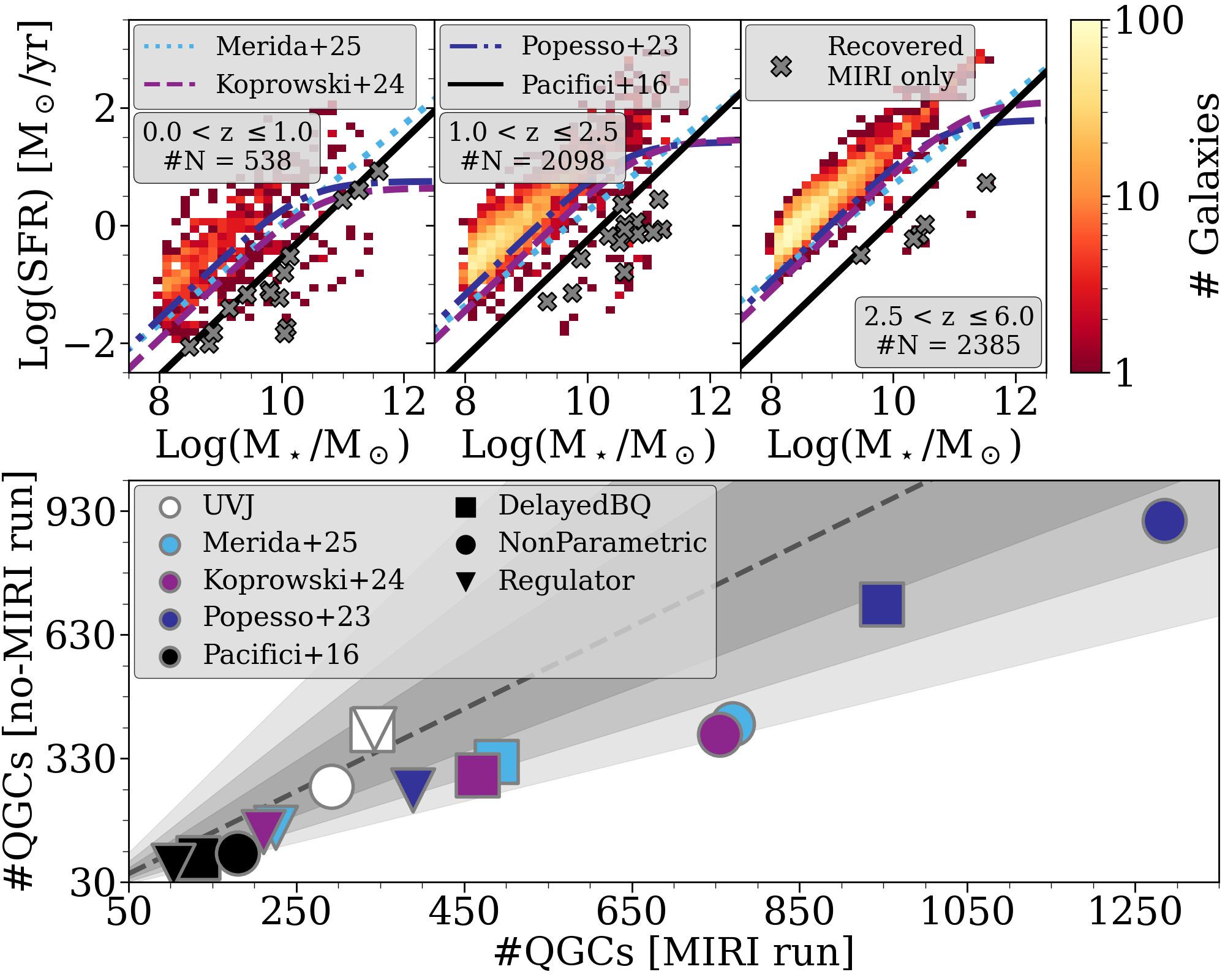}
\caption{
Top: SFR-M$_\star$ plane for different redshifts with galaxies from the final sample from \textit{MIRI run} with Regulator SFH. 
The different lines present quiescence selection from literature: black -- sSFR criterion, dark blue -- \cite{Popesso23}, violet -- \cite{Koprowski24}, light blue -- \cite{Merida25}. The grey crosses represent QGCs  selected by sSFR criterion in Regulator, \textit{MIRI run}, which were not recovered in \textit{no-MIRI run}. 
Bottom:
Comparison of the QGCs sample size depending on the selection criterion and SFH model for the \textit{MIRI} and \textit{no-MIRI runs}. The colours of the points represent the MSs and are the same as at top panel. The markers represent SFH model: triangle -- Regulator, circles -- NonParametric, squares -- DelayedBQ. The dashed line shows one-to-one relation. With the shaded regions we mark 30\%, 60\% and 100\% relative distance from the one-to-one relation. 
}
\label{fig:nQGs}
\end{figure}

To select QGCs with MS, we apply the same widely used MS offset to all scaling relations: ${\log(\rm{SFR/SFR}_{\rm{MS}})<-0.6}$ \citep[e.g.][]{Elbaz18,Donevski20, Lisiecki23}, where SFR$_{\rm{MS}}$ is the SFR predicted by given MS according to the stellar mass and redshift of the galaxy.
The visual interpretation of MS based quiescence thresholds are shown at Fig.~\ref{fig:nQGs}.

The choice of MS, while maintaining the SFH model, influences the number of selected QGCs up to factor of $\sim$2. 
This is expected since each MS is defined by the use of different samples, where even a selection of star-forming galaxies strongly influences the MS shape \citep[][]{Pearson23}.
We find that the choice of SFH influences the number of selected QGCs stronger by up to a factor of $\sim$3.5 when comparing NonParametric with Regulator models(see Fig.~\ref{fig:nQGs}). 
Comparing the sample size of QGCs selected with Regulator vs DelayedBQ, we find that they are different by a factor of $\sim$2-2.5. 
This is still larger compared to the differences between the estimation of physical properties with the use of the Regulator and the DelayedBQ model (see Fig.~\ref{fig:Physical_diff}).

The influence of the SFH model is independent of the inclusion of MIRI points. 
The NonParametric model results in the largest sample, both in \textit{MIRI} and \textit{no-MIRI runs}.
This is due to the early  stellar mass assembly achieved with this model, resulting in faster quenching \citep[][]{Annunziatella25}.
In contrast, the Regulator model consistently results in the least number of QGCs.
This is an effect of residual star formation that is always present with the use of this SFH model, as the gas reservoir can never be completely empty.

Finally, by adding MIR information to the SED fitting, we systematically reach a higher number of QGCs, by a factor of $\sim$1.4-2. 
This is in line with the fact that MIR allows better constraining of A$_V$ and SFR at the same time, which was explained in a previous section.
MIRI measurements puts soft prior on dust emission part of SED and allows for better interpretation of UV slope, lowering the SFR.
We mark the QGCs unselected with \textit{no-MIRI run} in the top panels of Fig.~\ref{fig:nQGs}.
It shows that MIR data influences the selection across all M$_\star$ and SFR.

\subsubsection{sSFR selection}
We test the criterion proposed by \cite{Pacifici16}.
According to this criterion, the galaxy is considered quenched when the specific SFR (sSFR; $\textrm{sSFR}\equiv \textrm{SFR/M}_*$) is lower than $0.2/\tau(z)$, where $\tau(z)$ is the age of the Universe at given redshift, in our case the observed redshift of the galaxy. 
We refer to this quiescence criterion as an sSFR criterion hereafter.
The thresholds originate from the observed evolution of the normalisation of the MS \citep[i.e.][]{Fumagalli14}.
It has been calibrated up to $z\sim2$ \citep[][]{Pacifici16,Montero19}, and have been validated in SIMBA cosmological simulation studies \citep[i.e.][]{Appleby20, Lorenzon25}.

\begin{figure*}[ht]
\centering
\includegraphics[width = 0.95\textwidth]{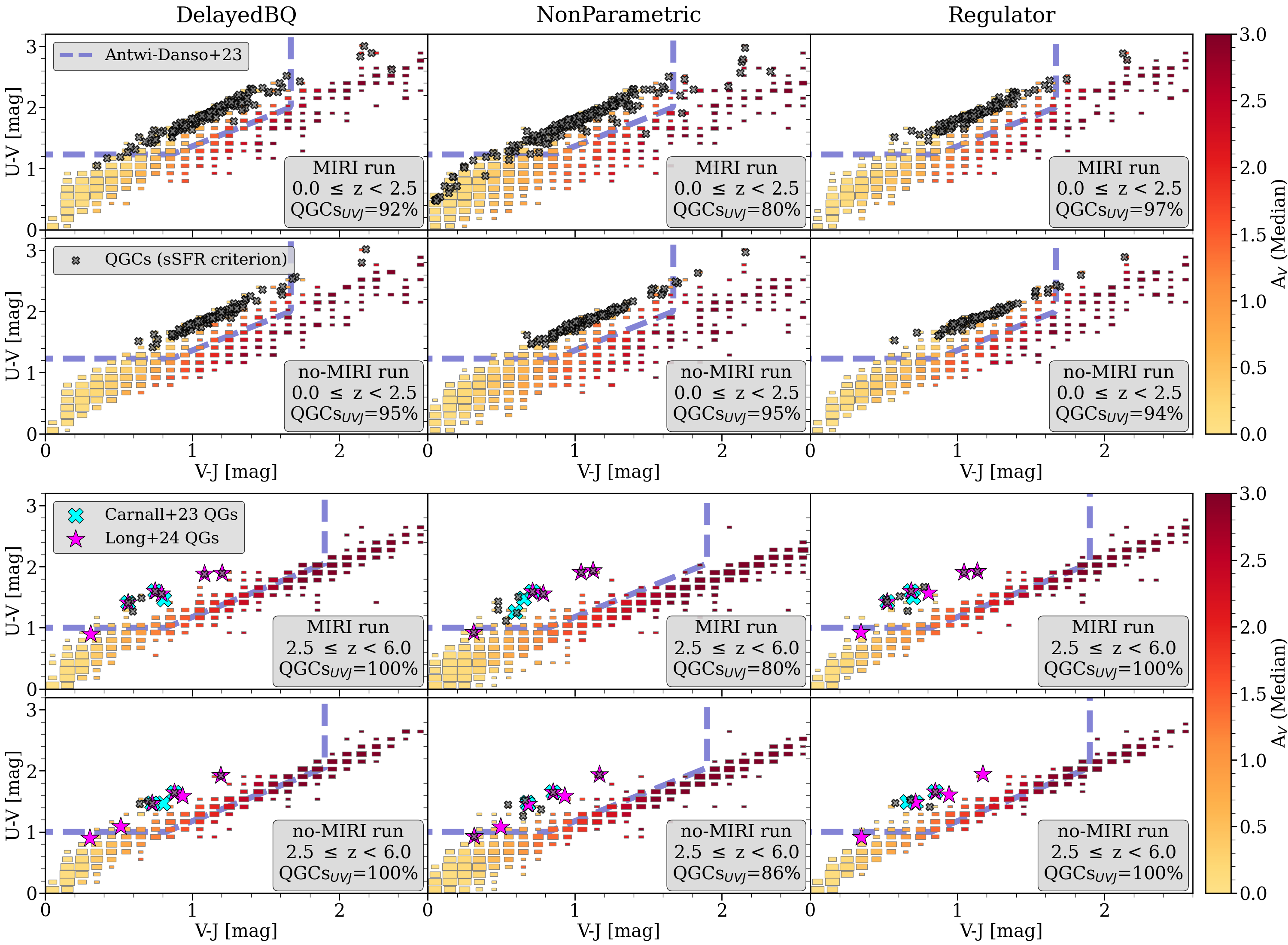}
\caption{Comparison of UVJ diagrams for galaxies with $z<2.5$ (top 6 panels) and $2.5\leq z<6$ (bottom 6 panels) for mass-complete sample (M$_*\geq10^8$M$_\odot$) from different runs. The squares are sized logarithmically according to the number of galaxies in the bin. We colour-code the median attenuation in V-band (A$_V$) as the histogram in the background. The grey crosses represent the QGCs selected with sSFR criterion. The blue dashed line shows the redshift-dependent criterion for quiescence by \cite{Antwi-Danso23}. We mark the fraction of QGCs (sSFR criterion) caught by UVJ criterion in the lower right corner of each panel. The top panels show the results from the \textit{MIRI run} and the bottom panels show the results from the \textit{no-MIRI run}. The columns are related to the SFH model used in the run, from left: DelayedBQ, NonParametric, and Regulator. Additionally, we include galaxies for which we are able to recover as QGs using \cite{Koprowski24} MS offset criteria: with cyan crosses QGs from \cite{Carnall23b} and with magenta stars QGs from \cite{Long24}. For details check App.~\ref{APP:UVJ}.}
\label{fig:UVJ}
\end{figure*}

Although the sSFR-based criterion is rooted in the MS framework, we find it more conservative, yielding the smallest quiescent sample (Fig.~\ref{fig:nQGs}). The impact of MIRI photometry persists: including MIRI in the SED fits increases the number of QGCs by a factor of $\sim\!1.6$. Differences among SFH parameterisations also remain at the $\sim1.3$-$1.8$ level; we find that the Regulator model returns the fewest QGCs, whereas the NonParametric model returns the most.

\subsubsection{UVJ selection}

It has been shown that the UVJ diagram can separate galaxies according to their sSFR and dust attenuation \citep[][]{Williams09,Whitaker11, Patel12, Donnari19, Antwi-Danso23, Valentino23}.
Although this empirical finding is consistent with properties estimated with standard attenuation curves and parametric SFH models, UVJ selection may be not representative for all QGCs across masses and redshifts \citep[][]{Leja19UVJ, Akins22,Antwi-Danso23, Long24, Gebek25}. 
We show the UVJ colour-colour diagrams for the final sample (mass complete, $z<6$) in Fig.~\ref{fig:UVJ}.

We construct the UVJ diagrams using the CIGALE results from each of the runs.
They reflect the known correlation between the rest-frame colours and A$_V$ (increase with higher colours values) and sSFR \citep[decrease with lower V-J and higher U-V colours, see][]{Williams09, Martis19, Akins22}.
However, there are significant differences among the selected samples of QGCs. 
We test the bias between the UVJ, MS offset, and sSFR criteria.
The results are presented in Tab.~\ref{TAB:UVJ}.

\begin{table}[ht] 
\centering
\caption{Comparison of selected QGCs in final sample and low-$z$ sample.}
    \begin{tabular}{l r r}
    Run &  \#QGCs ($z<2.5$) & Out ($z<2.5$)\\
    \hline
    \multicolumn{3}{c}{sSFR criterion; \cite{Pacifici16}}\\
    \hline
    \hline
    DelayedBQ \phantom{000}\textit{no-MIRI} & 86 \phantom{0}(81) & \phantom{00}4 \phantom{00}(4)\\
    DelayedBQ \phantom{000}\textit{MIRI} & 128 (120) & \phantom{00}9 \phantom{00}(9)\\
    \hline
    NonParametric \textit{no-MIRI} & \phantom{0}93 \phantom{0}(86) & \phantom{00}5 \phantom{00}(4)\\
    NonParametric \textit{MIRI} & 174 (164)& \phantom{0}34 \phantom{0}(32)\\
    \hline
    Regulator \phantom{0000}\textit{no-MIRI} & 67 \phantom{0}(63) & \phantom{00}4 \phantom{00}(4)\\
    Regulator \phantom{0000}\textit{MIRI} & 100 \phantom{0}(92) & \phantom{00}3 \phantom{00}(3)\\
    \hline
    \multicolumn{3}{c}{MS offset; \cite{Koprowski24}}\\
    \hline
    \hline
    DelayedBQ \phantom{000}\textit{no-MIRI} & 270 (231) & 120 \phantom{0}(89)\\
    DelayedBQ \phantom{000}\textit{MIRI} & 445 (395) & 245 (204)\\
    \hline
    NonParametric \textit{no-MIRI} & 371 (300)& 371 (150)\\
    NonParametric \textit{MIRI} & 740 (661)& 740 (443)\\
    \hline
    Regulator \phantom{0000}\textit{no-MIRI} & 141 (127) & \phantom{0}13 \phantom{00}(7)\\
    Regulator \phantom{0000}\textit{MIRI} & 200 (179) & \phantom{0}28 \phantom{0}(20)\\
    \hline
\end{tabular}
\tablefoot{The table is split in two sections: top section defines QGCs by sSFR criterion; the bottom section defines QGCs using \cite{Koprowski24} MS distance. The first column describes the run, the second one shows the number of all QGCs in the run. The number in the bracket shows the number of QGCs at $z<2.5$. The third columns shows the number of QGCs that are UVJ unselected by quiescence criterion proposed by \citealt{Antwi-Danso23}. The number in the bracket shows the number of QGCs unselected by UVJ criterion at $z<2.5$.}
\label{TAB:UVJ}
\end{table}

Up to$\sim20$\% QGCs (selected with sSFR criterion) are inconsistent with being a SFG according to the UVJ selection (UVJ unselected, hereafter) proposed by \cite{Antwi-Danso23}.
With DelayedBQ and NonParametric SFH models, MIRI points enlarge the number of UVJ unselected, and the opposite effect is observed with Regulator model. 
This is true for both the entire sample and the sample restricted to $z<2.5$ (see Tab.~\ref{TAB:UVJ}).

When selecting QGCs based on their MS distance \citep[as defined in][]{Koprowski24}, the fraction of QGCs that are not selected by UVJ criterion is strongly biased by both SFH and MIRI.
Firstly, MIRI points tend to always increase the number of QGCs not selected by UVJ, exposing that UVJ does not break completely the degeneracy of sSFH-A$_V$.
Secondly, the fraction of UVJ unselected QGCs is always slightly lower in the sample restricted to $z<2.5$ than in the entire sample.
This is in agreement with recent studies showing that UVJ can not separate QGCs from the other galaxies for high-$z$ galaxies ($z>3$) due to still relatively young stellar populations \citep[][]{Antwi-Danso23,Lovell23, Valentino23, Long24}.
Finally, the number of QGCs not selected by UVJ is strongly dependent on SFH and is the highest for NonParametric model, reaching 69\%. 
The Regulator model is the most consistent with \cite{Schreiber15} UVJ criterion, but the fraction of QGCs not selected by UVJ can still reach 14\%. 

We confirm that our selection methodology successfully recovers the known and spectroscopically confirmed QGs at high-$z$ from \cite{Carnall23b} and the QGs studied in \cite{Long24}.
For a detailed description, see the App.~\ref{APP:UVJ}.
In summary, we find that three sources from \cite{Carnall23b} and 8 sources from \cite{Long24} are present in our final sample (the rest were not in MIRI pointings). 
We recover similar M$_\star$ ($\pm0.3$ dex) and we are able to recover the quiescent nature of all three \cite{Carnall23b} galaxies and 5-6 (depending on the SFH model) out of nine \cite{Long24} galaxies using the \cite{Koprowski24} MS offset criterion. 
All recovered QGs are marked in Fig.~\ref{fig:UVJ}.

\begin{figure*}[ht]
\centering
\includegraphics[width = 0.98\textwidth]{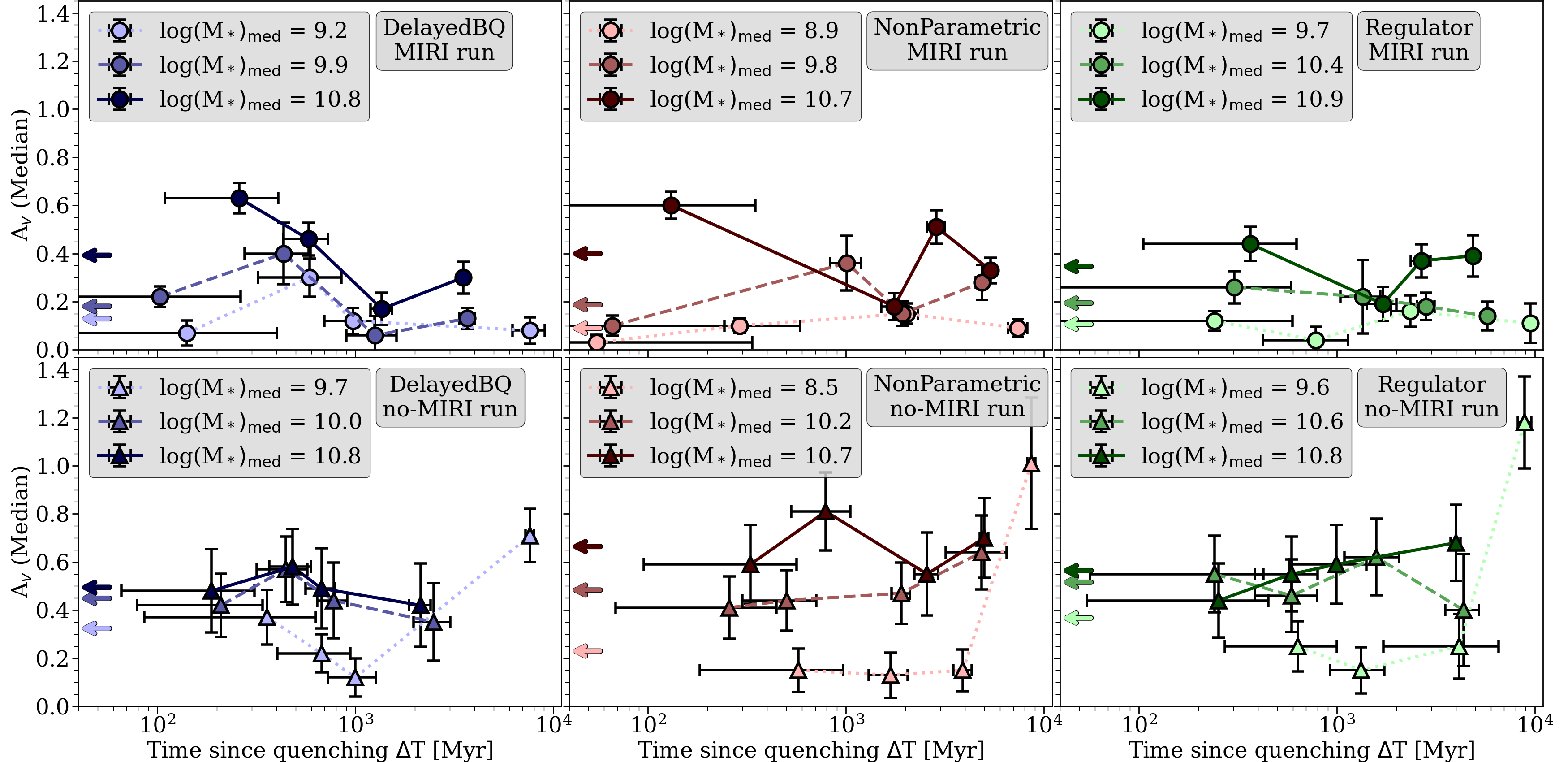}
\caption{Relation between time since quenching ($\Delta$T) and the A$_V$ of the mass-complete QGCs at $z\leq2.5$. The columns show different SFH models used in the run, from left DelayedBQ, NonParametric, and Regulator, while the rows show \textit{MIRI} vs \textit{no-MIRI runs}. The colours represent the median M$_\star$ of each bins. The arrows on the left of each panel show the weighted mean for the corresponding M$_\star$ bin.}
\label{fig:Av_T}
\end{figure*}

We find that the mean A$_V$ increases with M$_\star$ for star-forming galaxies (non-QGCs) in all runs in both redshift bins.
Our results in redshift $z<2.5$ align perfectly with the relation presented by \cite{vanderWel25}. 
The V-J colour increases with A$_V$ values, and low and intermediate-mass galaxies (M$_*<10^{10}$M$_\odot$) tend to have A$_V< 1$ mag.
For galaxies with M$_*>10^{11}$M$_\odot$ mean A$_V$ reaches $\sim$3 mag.
Although this trend is also visible for QGCs in all runs, the A$_V$ increases only by 0.26, 0.3 and 0.25 mag within \textit{MIRI run} for DelayedBQ, NonParametric and Regulator model, respectively (see Fig.\ref{fig:Av_T}).

\subsection{Inferring quenching-related parameters from SFH}\label{Sec:InferringQuenching}

Finally, we provide insight into the quenching properties and timescales that are recovered by our photometry-fitted SFH models.
For this purpose, we prepare and study sSFH for each QGC selected by the sSFR criterion.
The description of sSFH calculation can be found in App.~\ref{APP:sSFHs}.
In short, we recreate the M$_\star$ growth curve according to SFH and divide the SFH by the M$_\star$ growth curve to obtain sSFR in each time step.
Here we use low-$z$ part of our sample ($z\leq2.5$), as the sSFR-based criterion was not calibrated for higher redshifts, and MIRI fluxes directly probe the warm dust emission within this redshift range.

For each QGC, we define three key quenching timescales to quantify the temporal changes of SFR: (1) quenching moment $\tau_q$; (2) quenching time T$_q$; (3) time since quenching $\Delta$T.
For a visual interpretation, see Fig.~\ref{fig:examplesSFH}. 
It is worth mentioning that the examples in the figure are not representative of the low- and high-mass populations.
The $\tau_q$ is the cosmic time when the galaxy is considered quenched (the sSFR of the galaxy goes below $0.2/\tau(z)$, where $\tau(z)$ is the cosmic time at a given redshift).
To account for stochastic changes in SFR changes, we follow \cite{Lorenzon25} and require that the sSFR remains below the threshold for a minimum time of $0.2 \times\tau(z_q)$, where $z_q$ is the redshift to fall below the threshold.
The T$_q$ is the time the galaxy spends between two thresholds defined as $1/\tau(z)$ and $0.2/\tau(z)$ (see examples at Fig.~\ref{APP:simulatedGal}).
In other words, T$_q$ describes how rapid the quenching process was.
The $\Delta$T is the look-back time since the quenching moment ($\Delta$T~$\equiv\tau(z_o) - \tau_q$, where $z_o$ is the observed redshift of the galaxy).

We study the consistency amongst quenching parameters defined above in the same way as in Sec.~\ref{sec:imapctSFH} and App.~\ref{APP:stabilityOfPhysics}.
We use the five best fits per galaxy (see Sec.~\ref{sec:SEDmodeling}).
We perform stability analysis for the Regulator model only, since it is free of instant quenching processes.
The 90th percentile, P$_{90}$ of the CV distribution of quenching time ($\tau_q$) is at $\sim$ 43\%. 
The influence of redshift on the stability of $\tau_q$ is negligible.
Thus, the $\tau_q$ recovered by individual runs can vary at the level of <45\%.
The P$_{90}$ of the T$_q$  CV distribution is at $\sim$199\%. 
Thus, T$_q$ does not converge to the same value and the estimate is not stable.

By comparing the recovered timescales to those found in cosmological simulations for objects selected the same way, we confirm their physical meaning.
For this purpose, we use an exemplary simulated galaxy from SIMBA (for details, see App.~\ref{APP:simulatedGal}).
We report that the value of $\tau_q$ is consistent within 2$\sigma$ (with $\sigma\sim$8\% of the value) compared to the simulated SFH.
Although the value of T$_q$ from the best fit is consistent with the one obtained directly from the simulation, it again does not converge to the same value as $\sigma \sim$ 67\%. 
Thus, due to the large scatter in the observational sample and in the SED modelling of the simulated galaxy, we do not recommend using T$_q$ with photometry-only studies. 

\subsection{Are QGCs dust attenuated?}

In the \textit{MIRI} run, using the sSFR criterion, we find 16 (13\%), 24 (14\%), and 4 (4\%) QGCs with $A_V>1$ for the DelayedBQ, NonParametric, and Regulator SFH models, respectively; percentages are relative to the total number of QGCs per model.
In Fig.~\ref{fig:examplesSFH}, we present the ranges for observational fluxes depending on attenuation, as well as the number of sources in the four attenuation bins for \textit{MIRI run} with Regulation SFH.
In the \textit{no–MIRI} run, for the same combination of SFH models, we find 9 (10\%), 10 (11\%), and 3 (9\%) QGCs with $A_V>1$.
It is clear that both the absolute counts and the fractions of dusty QGCs are lower without MIRI, underscoring the importance of mid-IR constraints for robust $A_V$ estimates and sample classification.
Obtained numbers also indicate that QGs can maintain a significant amount of dust after quenching. This points to mechanisms that can prevent or even rebuild the dust reservoir in quenched objects. 
In literature there is a growing number of studies identifying dusty QGCs across redshifts\citep[see ][]{Donevski23, Setton24, Siegel24, Bevacqua25, Lorenzon25b}, but the mechanisms behind their dusty nature are not yet fully understood \citep[][]{Lorenzon25}.

In Fig.~\ref{fig:Av_T} we show the evolution of A$_V$ as a function of time since quenching for selected QGCs at $z\leq2.5$ and considering all combinations of SFHs and inclusion/exclusion of MIRI data.
We study a mass-complete sample.
The QGCs are divided into three M$_\star$ bins of similar size.
In each M$_\star$ bin, the QGCs are divided into four $\Delta$T sub-bins of similar size (the number of QGCs in sub-bins is always $\geq$6). 
The ranges of M$_\star$ and $\Delta$T for each sub-bin are presented in Tab.~\ref{TAB:deltaTbins}.
In each sub-bin, we calculate the median value of $\Delta$T and A$_V$.
The error bars represent the standard deviation of the bootstrap analysis, a resampling method to estimate the statistical uncertainties of the measurements combined with intrinsic uncertainties.
We resampled our data set for 1\,000 random samples with three random objects removed from the entire sample of mass-complete QGCs. 

Comparing the \textit{MIRI} and \textit{no-MIRI runs} one can see, in Fig.~\ref{fig:Av_T}, that the addition of MIRI points and dust emission models reduces the median A$_V$ in all M$_\star$ bins.
We discuss the constraining of A$_V$ according to the CIGALE mock analysis in App.~\ref{APP:miriAttenuation}.
Despite the negligible change in A$_V$ uncertainties, consistency amongst different SFH is stable across the probed $\Delta$T ranges.
This decreased A$_V$ is more prominent towards larger $\Delta$T and lower M$_\star$.
We see a dramatic decrease in A$_V$ (0.5-1.2 mag) when MIRI data is present for the largest $\Delta$T and least massive QGCs (M$_*\sim10^9$M$_\odot$).
Even if obtaining proper ages and A$_V$ is uncertain without spectroscopy \citep[e.g.][]{Nersesian25}, we can use rich photometry to constrain them. 
Thus, MIRI data help break the dust-age degeneracy.

Interestingly, as shown with the arrows at left of each panel of Fig.~\ref{fig:Av_T}, the most massive bin always has the highest attenuation. 
A similar relation of A$_V$ as a function of M$_\star$ was reported by \cite{Gebek25, Maheson25} and \cite{vanderWel25}.
We suspect that this is an effect of a patchy dust distribution, or dusty cores, rather than extreme dustiness \citep[][]{Miller22, Setton24, Siegel24}.
Although this is true for weighted mean values, we do not see a direct evolution of A$_V$ with $\Delta$T.

QGCs with lower M$_\star$ tend to have a larger span of $\Delta$T.
For DelayedBQ and Regulator runs, the distributions of $\Delta$T in the least massive bin (M$_*\sim10^9$M$_\odot$) is twice as wide as for the most massive bin (M$_*\sim10^{11}$M$_\odot$).
For NonParametric runs, it has the same ratio only for the \textit{no-MIRI run}, while for the \textit{MIRI run} it is still visible, but reaches only $\sim$25\%.
This indicates distinct evolutionary paths for high- and low-mass QGCs.
A similar finding was reported by \cite{Cutler24}.
It may be related to the quenching mechanism, as reported for low-$z$ galaxies \citep[][]{Donnari19}.

Our analysis of all three \textit{MIRI runs} reveals a consistent trend for sustaining substantial A$_V$ first $\sim1$~Gyr after quenching.
This is in line with the findings and predictions of \cite{Donevski23}, \cite{Lorenzon25} and \cite{Lorenzon25b}.

We report that QGCs with $\Delta$T$>1$~Gyr (mature QGCs, hereafter) are the most diverse.
Firstly, most mature QGCs maintain low attenuation, A$_V<1$ mag.
This indicates that dust production is inefficient in typical QGCs or that strong destruction processes dominate the post-quenching phase.
Nevertheless, the most attenuated QGCs in all runs are usually mature as well.
All QGCs with A$_V>1$ mag within Regulator \textit{MIRI run} are mature.
For the NonParametric \textit{MIRI run}, we find that 55\% of QGCs with A$_V>1$ mag are mature, however, this fraction grows to 77\% for A$_V>1.2$ mag.
Only the DelayedBQ \textit{MIRI run} results in a lower fraction of mature QGCs with A$_V>1$ mag, reaching 30\%. 
These mature and attenuated QGCs had more than 1 Gyr to accumulate dust, which implies that the mechanism responsible for sustaining the dust operated on long timescales rather than rapidly.
This diversity may point to an additional mechanism that enriches ISM with dust proposed in some studies \citep[e.g. dust re-growth][]{Donevski23, Lorenzon25} or the addition of asymptotic giant branch (AGB) stars \citep[][]{Michalowski19, Bevacqua25}.

\section{Summary and conclusions}
We exploited the unprecedented depth of $\sim$29 mag of the JWST data and the wide optical-to-MIR wavelength coverage in the CEERS field to perform a comprehensive comparison of factors affecting the photometric selection of QGCs.
Combining HST/JWST fluxes from the ASTRODEEP-JWST catalogue over the CEERS field \citep[][]{Merlin24} with our analysis of MIRI detections and observational depths, we built a multi-wavelength (optical-MIR) catalogue of galaxies below $z<6$, including MIRI upper limits.
In total, we studied a mass complete sample of $\sim$ 5\,000 galaxies within the MIRI footprint.
We have investigated how the selection of QGCs and their physical properties are impacted due to different SFH choices, including NonParametric and Regulator stochastic models that we implemented within the CIGALE code.

Our main results can be summarised as follows:
\begin{itemize}
    \item We constructed the largest so far catalogue of MIRI observed QGCs. It consists of between 103 and 180 galaxies when the most restrictive QGC selection criterion \citep[sSFR, ][]{Pacifici16} is used; the factor of $\sim$2 difference in sample size depends on the SFH model used. When the MS distance (in particular, the \citealt{Koprowski24} MS) is used as the selection criterion, the corresponding number of selected QGCs increases to 200-743.
    \item Inclusion of MIRI points and dust emission models in the SED modelling is crucial for SFR estimation. It increases the number of QGCs selected by the factor $1.4-2$, independently of the selection criteria and the SFH model used.
    \item Different SFH models lead to large variations in number of selected QGCs, up to a factor of $\sim$3.5. The NonParametric model selects the largest samples, while the Regulator model yields the smallest.
    \item A significant fraction of QGCs defined either by sSFR or MS-offset are not recovered via UVJ diagrams. Incorporating MIR data and dust emission models reveals up to 69\% of QGCs missed by UVJ color-criterion alone.
    \item We report the importance of MIR observations in restricting A$_V$ in mature QGCs (with the time since quenching $\Delta$T$>1$ Gyr). 
    We did not find a clear correlation between A$_V$ and $\Delta$T. 
    However, we find that substantial dust attenuation is sustained after the first $\sim$1 Gyr after quenching. 
    This hints at an additional dust reprocessing mechanism being present in mature QGCs.
    \item Massive QGCs (M$_* \sim 10^{11}$ M$_\odot$) on average have A$_V$ greater by $\sim$0.3 mag as compared to the least massive QGCs in our sample (${\textrm{M}_*\sim 10^{9}\textrm{M}_\odot}$).
    This observation remains true for each SFH-data combination. 
    \item Photometry-only SED fitting codes, such as CIGALE, can recreate, to some extent, the SFH of the simulated galaxies. The precision of $\Delta$T reaches around $\sim$45\%, while T$_q$ does not converge to the same value. We urge caution on using $\tau_q$ inferred from the SFH modelled using photometric points only.
    \item Photometry alone can significantly constrain the evolution of galaxies, and MIRI is fundamental to obtain realistic A$_V$ and distinguish between QGs and SFGs.
\end{itemize}
Our results motivate QGC selection exploiting full photometric exploitation of NIRCam data with multiband MIRI imaging surveys in deep fields, including CEERS \citep[e.g.][]{Muzzin25}.

\begin{acknowledgements} 
We thank the anonymous referee for the careful reading of the manuscript and the useful comments improving the readability and clarifying the message.
We acknowledge support from the Polish National Science Center (NCN) grants: UMO-2023/49/N/ST9/00746, UMO- 2024/53/B/ST9/00230, 2023/50/E/ST9/00383, UMO-2020/39/D/ST9/00720, UMO-2023/51/D/ST9/00147, UMO-2020/38/E/ST9/00077, 2023/50/A/ST9/00579.
We acknowledge the support of the Polish Ministry of Education and Science through the grant PN/01/0034/2022 under `Perły Nauki' programme.
D.D. acknowledge support from the Polish National Agency for Academic Exchange
(Bekker grant BPN /BEK/2024/1/00029/DEC/1).
A.W.S.M. acknowledges the support of the Natural Sciences and Engineering Research Council of Canada (NSERC) through grant reference number RGPIN-2021-03046.
S.B. is supported by the ERC StG 101076080.
J. is funded by the European Union (MSCA  EDUCADO, GA 101119830 and WIDERA ExGal-Twin, GA 101158446).
C.B. acknowledges support through an Emmy Noether Grant of the German Research Foundation, a stipend by the Daimler and Benz Foundation and a Verbundforschung grant by the German Space Agency. 

\end{acknowledgements}

\bibliographystyle{aa}
\bibliography{main}

\begin{appendix} 
\section{Completeness analysis}\label{app:completeness}

\begin{figure*}[ht]
\centering
\includegraphics[width = 0.98\textwidth]{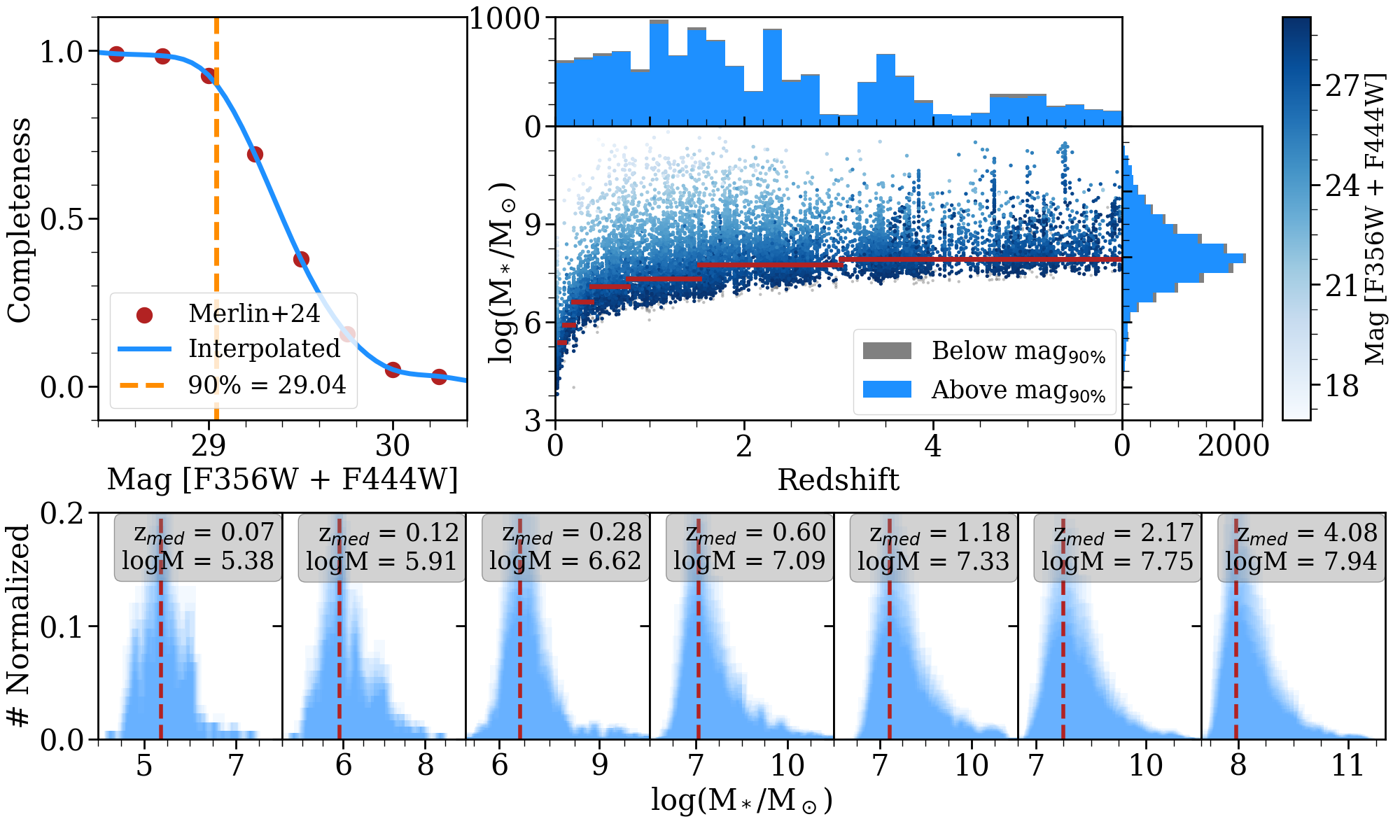}
\caption{Completeness analysis using the first method with Regulator \textit{MIRI run} results. Upper left panel presents the apparent magnitude completeness curve. The red points are values from \citep[][]{Merlin24}, the blue line is the interpolated curve and the orange dashed line is the value for 90\% completeness. Upper right panel presents the M$_\star$-redshift plane with histograms. In grey we mark galaxies below the mag$_{90\%}$ limit, while the white to blue colours represent the apparent magnitude of each galaxy. The red lines show the 90\% completeness for each redshift bin. The bottom panels show the histograms for each of the redshift bins. The estimated maximum of the distribution is shown with dashed red lines. The median redshift of the galaxies in bin is given a the top of each panel with the maximum of the distribution.}
\label{fig:massCompletenessAll}
\end{figure*}Impact of stochastic star-formation histories and dust information

To estimate the M$_\star$ completeness, we start with estimating the 90\% completeness in apparent magnitude (see Fig.~\ref{fig:massCompletenessAll}, upper left panel). 
For this, we use estimates from \cite{Merlin24} and we interpolate the curve using the cubic spline algorithm and find where the completeness curve falls below 0.9, mag$_{90\%} = 29.04$. 
Then we use two distinct methods to translate the apparent magnitude to M$_\star$.

For the first method, we select from our photometric catalogue of 12\,939 galaxies only the sources with apparent magnitude (F356W + F444W) above the 90\% completeness limit.
Then we divide the sample into seven logarithmic redshift bins between 0.05 and 6.
For each bin, we prepare 20 M$_\star$ histograms, each with a different number of bins between 15 and 35.
For each histogram, we find the M$_\star$ for which the distribution breaks (maximum of the distribution).
Finally, we take the maximum value of M$_\star$ from all histograms for a given redshift range as a 90\% completeness limit.
We report the log(M$_*/$M$_\odot$) limit for 90\% completeness to be 7.9, 7.8, and 7.9 for DelayedBQ, NonParametric and Regulator \textit{MIRI runs}, respectively.
This method is visualised in Fig.~\ref{fig:massCompletenessAll}.

\begin{figure}[ht]
\centering
\includegraphics[width = 0.48\textwidth]{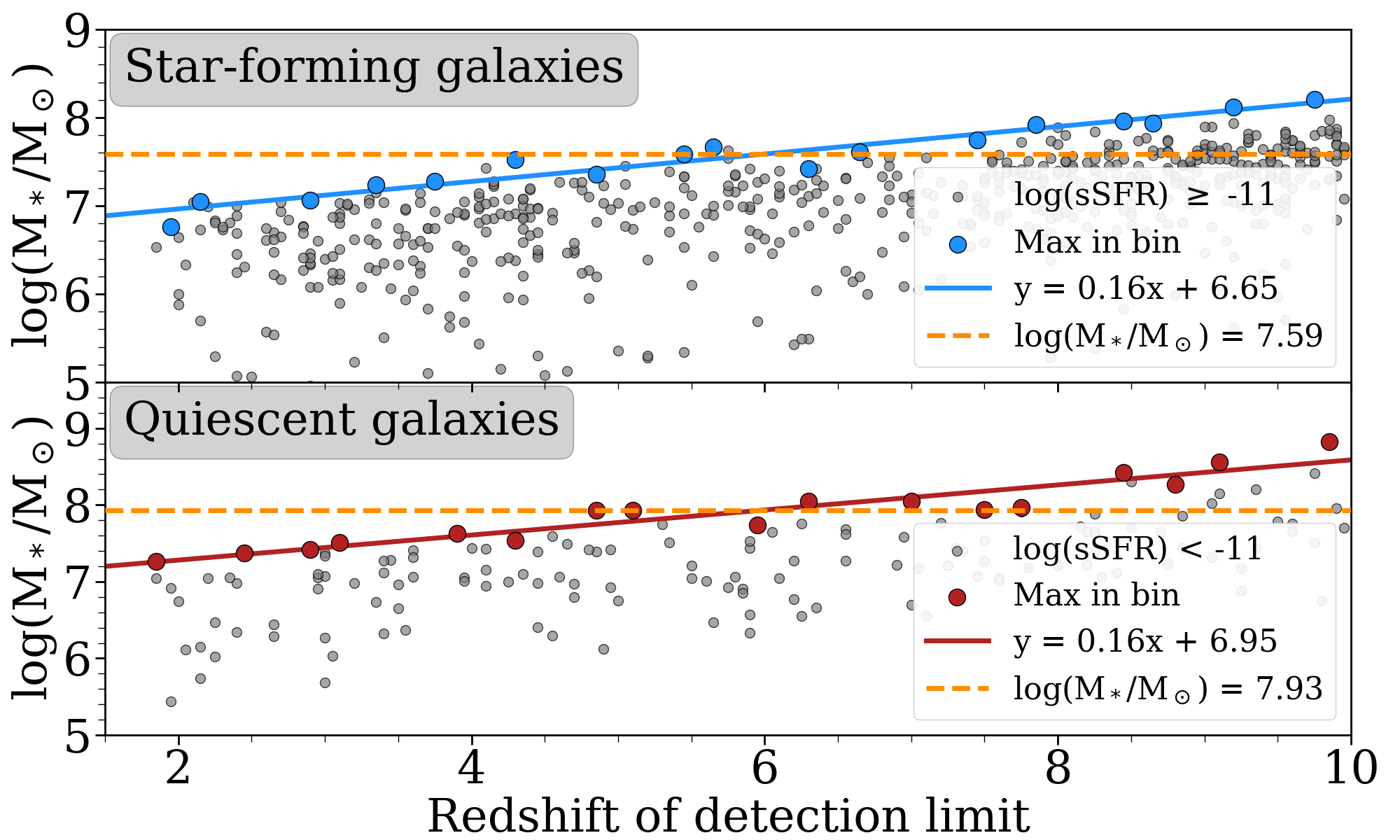}
\caption{M$_\star$ in function of maximum redshift for detection above 90\% completeness. The top panel presents the results for star-forming galaxies, while the bottom one presents the results for quiescent galaxies. The grey points are individual galaxies. The larger blue and red points present the largest M$_\star$ in each redshift bin. The blue and red solid lines show the linear fit, while the orange dashed line present the M$_\star$ at $z=6$ according to the fit.}
\label{fig:massCompletenessQG_SFR}
\end{figure}

For the second method, we utilise the stellar emission from the SEDs produced by CIGALE for all galaxies from our photometric catalogue of 12\,939 galaxies.
The SEDs are corrected to the rest-frame and redshifted for $z = 0 + 0.05n$, with $n = 1,2,3...200$ (up to $z = 10$).
Each of the redshifted SEDs is then convolved with a transmission curve of F356W and F444W to calculate the apparent magnitude. 
For each of the SEDs we find the maximum redshift for detection above 90\% completeness (the apparent magnitude of F356W + F444W greater than 29.04 mag).
We build two groups of galaxies:
\begin{equation}
\begin{aligned}
    &\textrm{Quiescent: }\log(\textrm{sSFR}[\textrm{M}_\odot/\textrm{yr}])<-11,\\
    &\textrm{Star-forming: }-9>\log(\textrm{sSFR}[\textrm{M}_\odot/\textrm{yr}])>-11.\\
\end{aligned}
\end{equation}
Then both groups are split into 0.5 wide redshift bins. 
We find the maximum M$_\star$ in each bin and fit a linear function to the results.
We report the completeness of 90\% in $z = 6$ in M$_*=10^{7.9}$ and $=10^{7.6}$ M$_\odot$ for the quiescent sample and the star-forming sample, respectively.
The results are presented in Fig.~\ref{fig:massCompletenessQG_SFR}.
Both methods are coherent and we can state that at $z=6$ our sample is 90\% complete for M$_*\simeq10^8$M$_\odot$.

\section{MIRI sources extraction}\label{APP:MIRI}

We use original images from the CEERS repository for public release 0.6\footnote{https://ceers.github.io/dr06.html}. 
Using Photutils.Background2D class \citep{Bradley23} with MedianBackground method and SigmaClip with sigma value of three, we calculate both background maps and RMS maps. 
Using SExtractor with default parameters and PSF Extractor \citep[PSFEx][]{Bertin11}, we extract empirical point spread functions (PSFs) using stars within the field of view for each MIRI filter used in CEERS. 

With extracted PSFs and GALFIT \citep{peng02}, we prepare model point-like sources with magnitude span 23--28 separated by 0.1 mag. 
We then insert 20 artificial, model sources into empty spaces in the images and run SExtractor, with extracted PSFs, to check which fraction of model sources is detectable.
As an empty space, we consider the point in the image with distance to all sources, real, and previously inserted, larger than twice the radius of the emission (parameter FLUX\_RADIUS for SExtractor sources or 15 pixels for the artificial sources).
As detection, we consider a source extracted by SExtractor with a distance from the inserted position less than 5 pixels ($\sim0.45$ arcsec).
We repeated the procedure 20 times per model source (in total, we insert each model 400 times).

\begin{table}[ht] 
\centering
\caption{Summary of SExtractor input.}
    \begin{tabular}{l c }
    Parameter& Value \\
    \hline
    DETECT\_MINAREA &  5\\
    DETECT\_THRESH   & 1.1\\
    ANALYSIS\_THRESH & 1.1\\
    DEBLEND\_NTHRESH & 64\\
    DEBLEND\_MINCONT & 0.0005\\     
    \hline
            
\end{tabular}
\label{TAB:SExtractorInp}
\end{table}

\begin{figure}[ht]
\centering
\includegraphics[width = 0.49\textwidth]{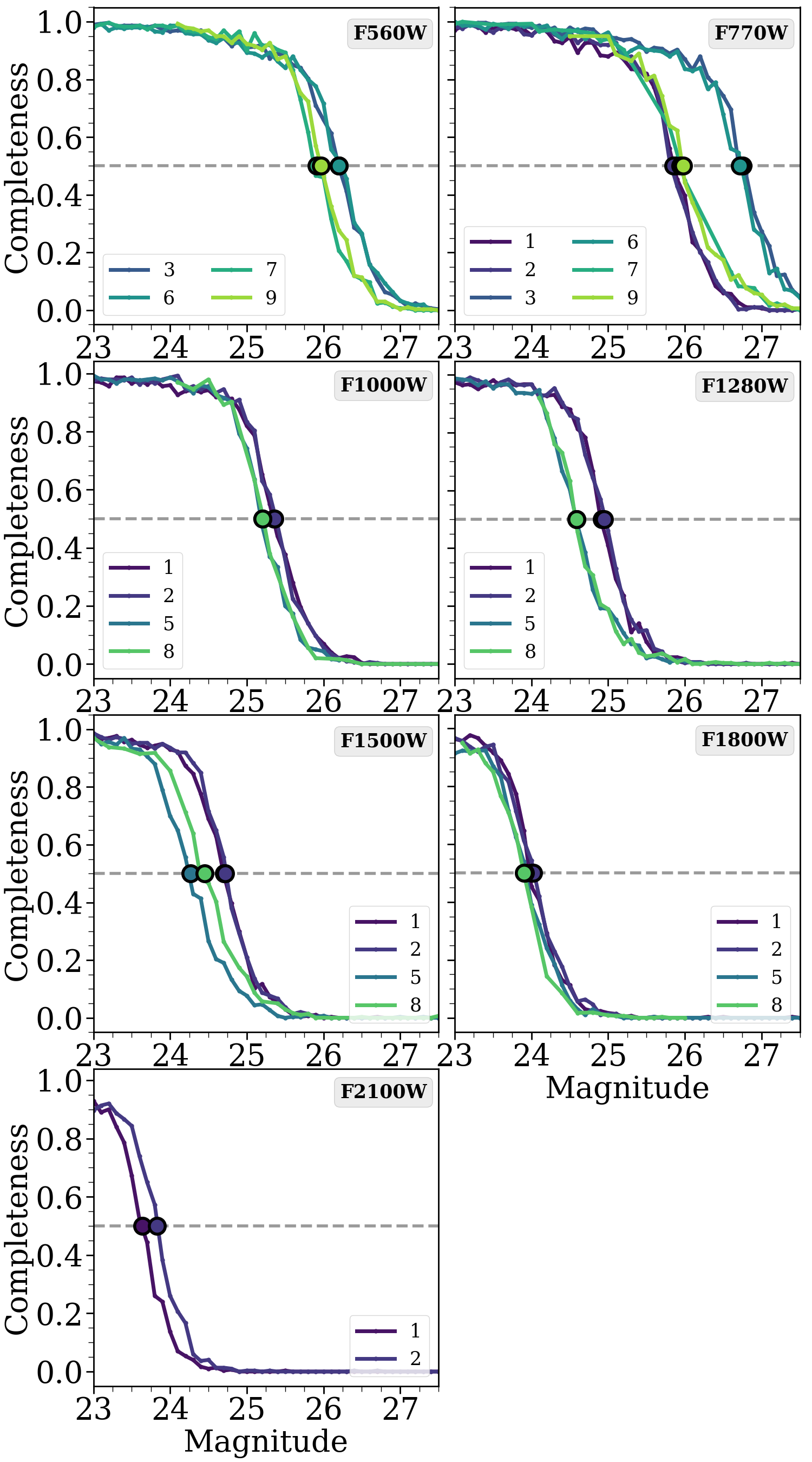}
\caption{Completeness curves for MIRI observations from the CEERS program. Each panel represents different band, which is written in the upper-right corner. Each coloured line represents MIRI pointing. The dashed, grey line marks completeness of 50\%, and the circles show the estimated 5$\sigma$ upper limit. }
\label{fig:completeness}
\end{figure}

We present the results in the form of completeness curves in Fig.~\ref{fig:completeness}, and as 5$\sigma$ upper limits in Tab.~\ref{TAB:MIRI_upperlimits}.
The upper limits are estimated by fitting linear function on the completeness points between $0.3 -0.7$ values and calculating the magnitude for which the completeness is $0.5$.
It is worth noticing, observations with the same filter (e.g. F770W) can have different depths depending on the pointing.
However, this is consistent with the integration time in each pointing. 

Finally, we perform the final SExtractor run with extracted PSFs, RMS maps as WEIGHT\_TYPE and other parameters specified in Tab.~\ref{TAB:SExtractorInp}. 
For each detection, we check if the ratio of flux (calculated from MAG\_AUTO) to error (calculated from MAGERR\_AUTO) is greater than 3.
If the fraction is lower, we report a 5 $\sigma$ upper limit as reported in Tab.~\ref{TAB:MIRI_upperlimits}.

\begin{table*}[ht] 
\centering
\caption{Summary of 5$\sigma$ detections in magnitudes through the MIRI imaging in CEERS.}
    \begin{tabular}{l c c c c c c c}
    Pointing& F560W & F770W & F1000W & F1280W & F1500W & F1800W & F2100W \\
    \hline
    MIRI1 & -- & 25.88 & 25.36 & 24.92 & 24.70 & 24.00 & 23.64\\
    MIRI2 & -- & 25.85 & 25.36 & 24.95 & 24.72 & 24.03 & 23.83\\
    MIRI3 & 26.20 & 26.75 & -- & -- & -- & -- & --\\
    MIRI5 & -- & -- & 25.20 & 24.59 & 24.26 & 23.92 & --\\
    MIRI6 & 26.20 & 26.72 & -- & -- & -- & -- & --\\
    MIRI7 & 25.91 & 25.95 & -- & -- & -- & -- & --\\
    MIRI8 & -- & -- & 25.21 & 24.59 & 24.45 & 23.90 & --\\
    MIRI9 & 25.97 & 25.94 & -- & -- & -- & -- & --\\
    \hline
            
\end{tabular}
\tablefoot{If given pointing was not observed within CEERS project with particular MIRI band, we put `--'.}
\label{TAB:MIRI_upperlimits}
\end{table*}

\section{CIGALE and implementation of stochastic SFHs}\label{APP:implementationSFH}

\subsection{Stochastic SFH I -- NonParametric}\label{SEC:nonparam}
The first stochastic SFH model that we test in our SED fitting process with CIGALE (Sec.~\ref{sec:SEDmodeling}) is referred to as NonParametric. 
We follow the approach described by \cite{Leja19} with prior continuity.
The continuity prior constrains the ratio of SFRs between the consecutive time-steps. 
This approach results in reduced burstiness, that is, fewer rapid changes in SFR in time.
The SFR changes follow the Student-t distribution, which is described by:
\begin{equation}\label{Eq:tstudent}
    \textrm{PDF}(x,\nu) = \frac{\Gamma\left(\frac{\nu+1}{2}\right)}{\sqrt{\nu\pi}\Gamma\left(\frac{\nu}{2}\right) } \left(1 +\frac{(x/\sigma_t)^2}{\nu}\right)^{-\frac{\nu + 1}{2}},
\end{equation}
where $\Gamma$ is the Gamma function, $\nu$ is the degree of freedom, and $\sigma_t$ is the scale factor.
Since with Student-t distribution, outliers are more likely to occur, it probes a broader range of SFH models than the Gaussian distribution. 
Using this distribution, we calculate the value of SFR as following:
\begin{equation}
    \textrm{SFR}(t_n) = \frac{\textrm{SFR}(t_{n-1})}{10^x},
\end{equation}
where $n$ is the index of the time step and $x$ is a random value from the Student-t distribution. 
For $n=1$ we set SFR$_n=1$ for numerical simplicity, since the total SFH is later normalised. 

We call this approach NonParametric; however, we are obliged to use several parameters.
In our implementation, we allow for five parameters (an exemplary set of parameters can be found in Tab.~\ref{TAB:CIGALE_inp}) in the non-parametric method.
The first one, \textit{age\_form}, is simply the look-back time of the formation of the observed galaxy, when the star formation starts.
The second, \textit{nModels}, controls how many SFHs should be prepared with posteriors of the Student-t distribution to test by CIGALE (each of them will be convolved with stellar population synthesis and enter a grid of models).
The next one \textit{scaleFactor}, controls $\sigma_t$ of the Student t distribution (see Eq.~\ref{Eq:tstudent}).
\cite{Ocvirk06} found that the evolutionary differences that can be observed and studied via SED in simple stellar populations are roughly proportional to their separation in logarithmic time.
To follow this result and build the SFH in logarithmic time bins, we use the last two parameters \textit{nLevels} and \textit{lastBin}.
The \textit{nLevels} controls the number of SFR changes, thus we divide \textit{age\_form} into \textit{nLevels} time steps.
The first time-step is defined as $1/\textit{nLevels}$ of \textit{age\_form}. 
The last time step is defined by \textit{lastBin}.
Finally, the rest of the time steps are equally spaced in logarithmic time (see Fig.~\ref{fig:examplesSFH} for visualisation).

Furthermore, we use the exact same priors for SFR changes when the only difference between the models is the \textit{age\_form}.
In this way, we minimise the risk of forcing CIGALE to choose the non-optimal SFH model due to random sampling.
In other words, if there is a run with more than one value for \textit{age\_form}, the priors will be sampled only once and the SFHs will be visually similar, just stretched. 
Following \cite{Leja19}, we adopt $\nu = 2$ (see Eq.~\ref{Eq:tstudent}), as it is based on the Illustris simulation results.

\subsection{Stochastic SFH II -- Regulator}\label{SEC:regulator}
The second stochastic SFH model that we test in our SED fitting process with CIGALE (Sec.~\ref{sec:SEDmodeling}) is defined as an Extended Regulator (Regulator hereafter). 
We follow the approach described in \cite{Iyer24}, first introduced in \cite{Tacchella20}.
The Regulator SFH model is based on the control of the SFR by the assigned gas mass reservoir.
The stochastic nature of the SFR is a result of 1) gas inflow/outflow rates, 2) gas cycling in equilibrium between atomic and molecular states, and 3) the formation, lifetime and disruption of giant molecular clouds (GMCs). 

As the gas mass budget may be affected by the processes mentioned above, we use the Regulator, which includes their effect.
We assume power laws to describe these effects \citep[e.g.][]{Iyer20, Tacchella20, Wang20} with power spectral density (PSD):
\begin{equation}
    \textrm{PSD} (f) = \frac{s^2}{1+4\pi^2\tau^2 f^2},
\end{equation}
where $f$ is the frequency of the process, $s^2$ is the absolute normalisation at $f=0$, and $\tau$ is the decorrelation time-scale of given process. 

Gas inflow and gas cycling in equilibrium are coupled, whereas the lifetimes of GMCs are independent of them \citep[][]{Tacchella20}. 
Thus, the total PSD will have two additive elements.
We calculate autocovariance function (ACF) from the total PSD using the Wiener–Khinchin theorem and describe the stochastic change of SFR using the following ACF:
\begin{equation}\label{EQ:ACFR}
    \textrm{ACF} = \sigma_{reg}^2\frac{\tau_{flow} e^{-|\tau|/\tau_{flow}} - \tau_{eq} e^{-|\tau|/\tau_{eq}}}{\tau_{flow} - \tau_{eq}} + \sigma_{dyn}^2 e^{-|\tau|/\tau_{dyn}},
\end{equation}
where $\sigma_{reg}$ is long-term variability related to inflow and equilibrium, $\sigma_{dyn}$ is the variability of the dynamical component, $\tau_{flow}$, $\tau_{eq}$, $\tau_{dyn}$ are timescales of gas inflow/outflow, equilibrium cycling and GMC's lifetime, respectively, and $\tau$ is the timescale associated with the frequency $f$.
For a detailed description of the derivation method of equation~\ref{EQ:ACFR}, we refer to \cite{Tacchella20} and \cite{Iyer24}. 

The ACF is now used as a physically motivated prior for stochastic SFH.
Following \cite{Iyer24}, we build SFH by assuming that at each time bin $t_n$ we can find SFR$_n$, where $n~=1,...,N$, which corresponds the multivariate normal distribution:
\begin{equation}
    \textrm{SFR}_n = \mathcal{N}\left( \mu_N, C(\tau) \right),
\end{equation}
where $\mu_N$ is the $N$-vector of mean values and $C(\tau)$ is the corresponding $N\times N$ covariance matrix. 
In the Regulator model, $C(\tau)$ is ACF$(\tau)$, for $\tau \equiv t - t'$, for $t, t' = t_1,...,t_N$, and $t_n$ are centres of the respective $N$ time bins.

In our implementation, we allow for eight parameters in the Regulator model (an exemplary set of parameters can be found in Table~\ref{TAB:CIGALE_inp}).
Namely, \textit{age\_form}, \textit{nModels}, which are exactly the same as in NonParametric model (see Sec.~\ref{SEC:nonparam}), \textit{nLevels} which controls in how many equal time bins \textit{age\_form} should be divided, and the equivalents of $\sigma_{reg}$, $\sigma_{dyn}$, $\tau_{eq}$, $\tau_{flow}$ and $\tau_{dyn}$ from Equation~\ref{EQ:ACFR}, respectively.
It is worth noticing that most stochastic processes are defined with the $\mu_N=0$.
Thus, following the implementation of \cite{Iyer24} or \cite{Wan24}, we also use the same mean vector.

\subsection{CIGALE input}\label{APP:Cigale}
Table \ref{TAB:CIGALE_inp} presents the CIGALE input used in this study.

\begin{table*}[!t] 
\centering
\caption{Summary of CIGALE input. If a parameter is not explicitly written, the initial values were kept.}
    \begin{tabular}{c c c}
    \hline
    \hline
    Parameter & Values & Description\\
    \hline
    \multicolumn{3}{c}{SFH ages of main stellar population}\\
    \hline
    age & 13 values with numpy.linspace from 500 to 13000 & For $0.0<z\leq0.7$ (2\,314 galaxies)\\
    age & 9 values with numpy.linspace from 500 to 7300   & For $0.7<z\leq1.5$ (2\,730 galaxies)\\
    age & 7 values with numpy.linspace from 500 to 4300   & For $1.5<z\leq2.6$ (3\,576 galaxies)\\
    age & 7 values with numpy.linspace from 250 to 2500   & For $2.6<z\leq4.0$ (2\,448 galaxies)\\
    age & 5 values with numpy.linspace from 150 to 1500   & For $4.0<z\leq6.0$ (1\,889 galaxies)\\
    \hline
    age\_bq & 5 values with numpy.logspace from 10 to 500  & For $0.0<z\leq2.6$ \\
    age\_bq & 5 values with numpy.logspace from 10 to 250  & For $2.6<z\leq4.0$ \\
    age\_bq & 4 values with numpy.logspace from 10 to 150& For $4.0<z\leq6.0$ \\
    \hline
    \multicolumn{3}{c}{SFH DelayedBQ}\\
    \hline
    age\_main & see age above &Age of the main stellar population\\
    age\_bq & see age\_bq above & Age of the late quenching event\\
    tau\_main & 500, 1600, 2750, 3900, 5000& e-folding time of the main stellar population\\
    r\_sfr& 0.01, 0.045, 0.22, 1, 3.& Ratio of SFR after and prior to quenching\\
    \hline
    \multicolumn{3}{c}{SFH NonParametric}\\
    \hline
    age\_form & see age above & Formation age of the galaxy\\
    nModels & 5$\times$200 (1000 in total)& Number of models\\
    nLevels & 8 & Number of time bins\\
    lastBin & 30 & Width of the most recent time bin \\
    \hline
    \multicolumn{3}{c}{SFH Regulator}\\
    \hline
    age\_form & see age above & Formation age of the galaxy\\
    nModels & 5$\times$200 (1000 in total)& Number of models\\
    nLevels & 300 & Number of time bins\\
    sigmaReg & 1& The long time variability of the gas\\
    tauEq & 1300& Time-scale of the equilibrium cycling\\
    tauFlow & 125 & Time-scale of the gas inflow\\
    sigmaDyn & 0.07 & The variability of GMCs\\
    tauDyn & 35 & Time-scale of the dynamical changes\\
    \hline
    \multicolumn{3}{c}{Stellar emission}\\
    \hline
    IMF & \cite{Chabrier03} & Initial mass function (IMF)\\
    \hline
    \multicolumn{3}{c}{Nebular emission}\\
    \hline
    logU & -3.0 & Ionisation parameter\\
    \hline
    \multicolumn{3}{c}{Dust attenuation}\\
    \hline
    Av\_ISM & 0, 0.1, 0.3, 0.5, 0.8, 1.2, 1.6, 2.1, 2.6, 3.2, 4, 5 & V-band attenuation in the interstellar medium\\
    slope\_ISM & -1, -1.7& Power law slope of the attenuation in the ISM.\\
    \hline
    \multicolumn{3}{c}{Dust emission}\\
    \hline
    qpah & 0.47, 1.77, 3.19 & Mass fraction of policyclic aromatic hydrocarbon (PAH)\\
    umin & 0.15, 2.0,10.0, 15.0, 25.0 & Minimum radiation field\\
    gamma & 0.001, 0.01, 0.1 & Fraction illuminated from Umin to Umax\\
    \hline  
\end{tabular}
\tablefoot{The age$_{\textrm{bq}}$ column refers to DelayedBQ (before quenching) CIGALE run.}
\label{TAB:CIGALE_inp}
\end{table*}

\section{Stability of physical properties}\label{APP:stabilityOfPhysics}

To ensure that splitting the run into five smaller runs does not influence the analysis, we have conducted one additional run with \textit{nLevels}$=1000$.
Comparing M$_*$, SFR, A$_V$ and age, we find that each of them is consistent despite the method chosen (one big run, and five smaller runs). 
The differences are approximately $\sim$10-20 times smaller than the error reported by CIGALE.
Additionally, the errors themselves are also consistent, within margin of $\sim$5\% of their value. 

The parametric SFH models, such as DelayedBQ always converge to the same value.
This does not have to be true for the stochastic SFH.
In order to probe the consistency of physical properties as outputs of our SED modelling with stochastic SFHs, we access the five (per stochastic SFH model) best resulting models for each galaxy within the final sample. 
We then check how stable the physical properties are between iteration with the same parameters, in other words, we wish to test to what level the model results (physical properties) are constrained.

\subsection{\textit{MIRI run}}\label{sec:MIRIRUNstability}
We first explore fitting the full SED, including the longer-wavelength MIRI data. 
The M$_\star$ is well preserved in the five best resulting models for each of the different stochastic SFHs. 
The 90th percentile (P$_{90}$) of the CV distributions is $\sim$1\% and $\sim$6\% for the Regulator and the NonParametric SFH models, respectively.
In other words, the Bayesian M$_\star$ converges to the same value up to $\sim$1\% or $\sim$6\% accuracy.
Similarly, ages of the galaxies tend to converge to the same values with P$_{90}$ of distributions $\sim$4\% and $\sim$6\% for Regulator and NonParametric SFH models, respectively.
While for the Regulator, SFR is stable throughout the iterations with P$_{90}$ at $\sim$3\%, the NonParametric distribution shows much larger scatter with P$_{90}$ at $\sim$17\%.
The strong variation of the SFR estimated with the NonParametric run is strongly biased due to lack of limit for the minimum SFR.
The CIGALE does not record the difference between SFR $=10^{-3}$ and SFR $=10^{-6}$, in both cases the young population of stars will be negligible.
Thus, the SFR of the passive galaxies in each iteration can be negligible but different. 
Finally, the CV distribution of A$_V$ shows a variation of 5\% and 8\% for the Regulator and the NonParametric model, respectively. 

\subsection{\textit{no-MIRI run}}
Now we explore fitting SED without longer-wavelength MIRI data.
The \textit{no-MIRI run} results have similar tendencies.
The P$_{90}$ of the M$_\star$ CV (relative standard deviation $\textrm{CV} = \sigma/\mu$) distributions is $\sim$1\% and $\sim$6\% for the Regulator and the NonParametric SFH models, respectively.
The ages tend to converge to the same values as P$_{90}$ of the distributions $\sim$5\% and $\sim$6\% for Regulator and NonParametric SFH models, respectively.
The CV distribution of the SFR within the Regulator has P$_{90}$ at $\sim$3\%, while within the NonParametric, the distribution has P$_{90}$ at $\sim$28\%.
The strong variation of the SFR estimated with the NonParametric run can be explained in the same way as in the previous section.
The attenuation varies 5\% and 7\% for the Regulator and the NonParametric model, respectively.

\subsection{Influence of MIRI data on photometric redshift}\label{APP:photoz}
We tested the influence of MIRI data on the estimation of photometric redshifts with CIGALE.
We prepared two additional, but simpler, runs with CIGALE.
Both runs use the same input parameters, differing only by the data used in the fitting: 1) the entire photometric data available in our catalogue; 2) HST and NIRCam data points only. 
We used 12\,939 galaxies from the MIRI footprint. 
We checked 30 different delayed SFHs (3 tau\_main and 10 age values), \cite{Chabrier03} IMF, dust attenuation by \cite{Charlot00} with 7 values of the Av\_ISM and \cite{Draine14} dust emission module with one single default output model. 
Finally, we allowed CIGALE to probe the redshift between 0.01 and 10 with a 0.003 redshift step.

\begin{figure}[ht]
\centering
\includegraphics[width = 0.48\textwidth]{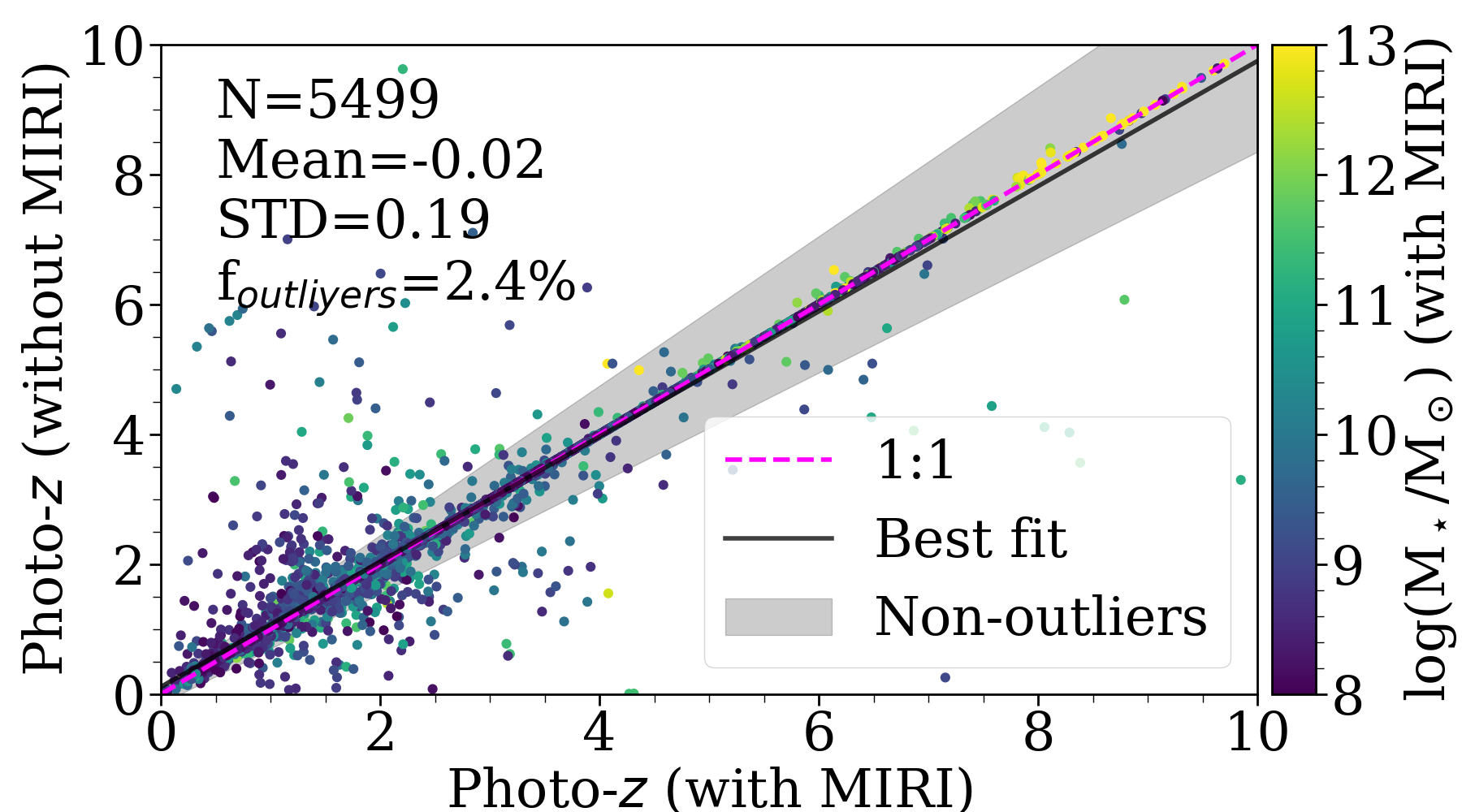}
\caption{Photo$-z$ estimation with and without MIRI data for galaxies with. The colour represent the stellar mass estimated in run with MIRI data. The magenta dashed line shows 1:1 relation, while the black solid line represent the best fit. The grey shaded region show the non-outliers region defined as ${|z_{withMIRI} - z_{withoutMIRI}|/(1+z_{withMIRI}) > 0.15}$.}
\label{fig:photozComparison}
\end{figure}

In Fig.\ref{fig:photozComparison}, we present the comparison of estimated photometric redshifts for galaxies with M$_\star>10^8$M$_\odot$ (according to run with MIRI). 
We report a great agreement between the estimates, with a mean value close to 0.
Additionally, the parentage of outliers, defined in the same way as in \cite{Merlin24}, as galaxies with ${|z_{withMIRI} - z_{withoutMIRI}|/(1+z_{withMIRI}) > 0.15}$, is also low, 2.4\%. 
Thus, the inclusion of MIRI photometry does not provide a systematic shift.
This is in agreement with a similar test provided by \cite{Wang25b}.

\subsection{Influence of MIRI data on attenuation}\label{APP:miriAttenuation}
We also test how does the A$_V$ change due to inclusion of MIRI data.
For this purpose we examine the change in A$_V$ through a mock analysis included in the CIGALE.
To produce the mock QGCs, the fluxes of each QGC from each run (see Sec.~\ref{SEC:QGsSelectionEff}) are varied within the error bars, building a perfect comparison sample of the same size as the observed one.
The best model is then fitted to each of the mock galaxies.
Comparison of real (not varied) results with mock ones shows how constrained the physical properties given are (M$_\star$, A$_V$, etc.).
We compare A$_V$ from the real run with the mock one for mass-complete QGCs within each run ($\Delta\textrm{A}_V\equiv \textrm{A}_{real} - \textrm{A}_{mock}$).
Here we switch to the 80th percentile (P$_{80}$) due to low statistics, in some bins the number of objects is less than 20.

Comparing \textit{MIRI} and \textit{no-MIRI runs}, we report that the P$_{80}$ of the $\Delta\textrm{A}_V$ distributions is consistently lower within the \textit{MIRI run}, by 9\%, 42\% and 33\% for DelayedBQ, NonParametric and Regulator, respectively.
Thus, the inclusion of MIRI data points in the SED fit helps to constrain the dust attenuation.
The $\Delta\textrm{A}_V$ is the largest in the NonParametric runs, reaching 0.26 (0.45) mag, while for DelayedBQ it reaches 0.22 (0.24) mag and for Regulator 0.18 (0.27) mag for \textit{MIRI} (\textit{no-MIRI}) \textit{run}. 
It is important to note that this is not constant in all $\Delta$T ranges.
This difference is more important for QGCs with larger $\Delta$T, and begins to differ significantly above $\Delta$T$\sim1$ Gyr.
In other words, to study A$_V$ in mature QGCs, the MIR fluxes are crucial.
This effect is less prominent in recently quenched galaxies.

\section{Extracting sSFH from CIGALE's results}\label{APP:sSFHs}
As discussed in Sec.~\ref{APP:stabilityOfPhysics}, the Bayesian set of results, such as M$_\star$ or age, seems to be consistent between runs.
However, as a CIGALE output, we get only the SFH corresponding to the best fit. 
Thus, we need to scale it to produce the Bayesian value of M$_\star$ in the Bayesian age. 
We explain the process in the following, but we also include the code to calculate sSFH in the GitHub project.\footnote{Ancillary files/Prepare\_sSFR.py}

First, we place the SFR in the cosmic-time array.
The CIGALE result SFH is a set of values, $T_{best}$, from 0 to the best fit age, age$_{best}$, of the galaxy in Myr and the corresponding table SFR$_{best}$.
Thus, we can place the galaxy in the cosmic time-frame for age$_{Bayes}$ via:
\begin{equation}
    T_{cosmic} = \textrm{age}_{Bayes}\left(\frac{T_{best}}{\textrm{age}_{best}} - 1\right) + t(z),
\end{equation}
where $t(z)$ is cosmic time at galaxy's observed $z$.
In this notation $T_{cosmic}$ is an array from the formation age, $t(z)-\textrm{age}_{bayes}$, up to $t(z)$. 
Finally, we can define $T_{cosmic,1Myr}$, which has the same age range but is equally spaced by 1 Myr.

We also have to scale the SFR table, SFR$_{best}$. 
However, due to the different ages range in $T_{cosmic}$, we first have to interpolate the SFR values for each corresponding time step in $T_{cosmic,1Myr}$.
In effect of interpolation, SFH now forms M$_{*,interp}$. 
Thus, we have to account for ratio of masses:
\begin{equation}
    \textrm{SFR}_{cosmic,1Myr} = \textrm{SFR}_{best} \frac{\textrm{M}_{*, Bayes}}{\textrm{M}_{*, interp}}, 
\end{equation}
where SFR$_{cosmic,1Myr}$ is the table of SFR corresponding to $T_{cosmic,1Myr}$.
The tables are now placed in cosmic time and form exactly M$_{*, Bayes}$.

To produce sSFH, we need to divide SFR$_{cosmic,1Myr}$ by the stellar mass of the galaxy.
To calculate the evolution of M$_\star$, we loop over each time step in $T_{cosmic,1Myr}$ and convolve the SFH up to the given time step with a single stellar population model used in our CIGALE run, calculating mass growth. 
We are aware this mass is only record of in situ formed stars and will differ when the stars were brought in through mergers. 
However, CIGALE does not include the history of mergers in the mass calculation \citep[][]{Iyer19}.
Thus, the in situ formed galaxy is the best we can achieve using CIGALE's SFH.

We must notice that, since we focus on reproducing Bayesian M$_\star$, the scaling can change the final SFR of the galaxy.
Thus, it is possible that the galaxy selected as QGC using CIGALE Bayesian output will not meet the same criteria after SFH scaling.
In our study, the difference in sample size never decreases by more than $\sim$3\% compared to selection with use of direct CIGALE SFR values.

\section{Validation of SFH with SIMBA galaxy}\label{APP:simulatedGal}

SIMBA \citep[][]{dave19} is a cutting-edge cosmological hydrodynamical simulation.
It evolves baryonic matter together with dark matter and is perfectly suited for studying dust-related properties because of the self-consistent evaluation of dust destruction and growth processes at each evolutionary step.
Furthermore, it implements a more realistic and careful treatment of the AGN feedback and the rich chemical evolution of ISM, compared to the previous iteration \citep[MUFASA, ][]{dave16}.
We make use of the full-physics high-resolution simulation (m25n512) consisting of a box side 25h$^{-1}$ Mpc containing 2$\times$512$^3$ particles. 
The entire run covers the redshift range 0 $<z<$ 249. 
For details of the simulation, we refer to \cite{dave19}.

In this study, we use a SIMBA simulated galaxy in order to compare direct, simulated SFH with the one estimated via SED fitting to check to what extent we can trust SFHs based purely on photometry.
For this purpose, we select the largest galaxy in the simulation at $z\sim1.5$, which is similar to the maximum of the redshift distribution in our observed sample.
We have coupled the SIMBA galaxy with the POWDERDAY radiative transfer code \citep[][]{Narayanan21}.
The selection of a massive galaxy was important from the perspective of the resolution of radiative transfer calculations. 

\begin{figure}[ht]
\centering
\includegraphics[width = 0.48\textwidth]{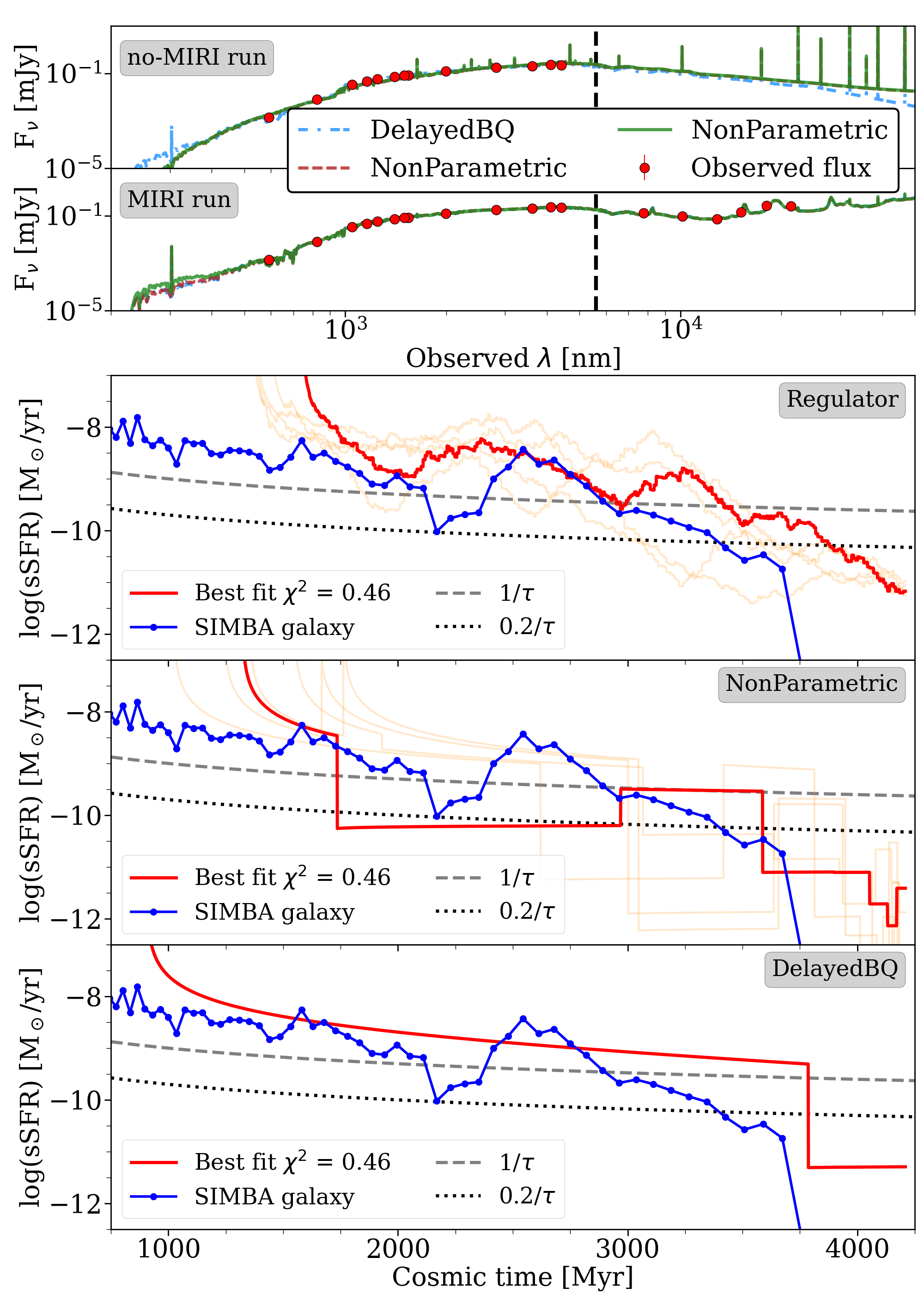}
\caption{Best SEDs and corresponding sSFHs of the SIMBA galaxy. The top figure shows comparison of SEDs from \textit{no-MIRI} vs \textit{MIRI} run. The observed fluxes, obtained by convolving filter with simulated SED, are marked with red circles. The colours and line styles of SEDs correspond to the used SFH. The black dashed line marks position of JWST/MIRI F560W filter.
The bottom figure compares the sSFHs directly from SIMBA and from CIGALE from \textit{MIRI run}. 
Starting from top panel: Regulator, NonParametric, DelayedBQ SFH. The blue line shows values of sSFH directly from SIMBA galaxy catalogue. The red line shows the best CIGALE fit. The orange lines in the background in Regulator and NonParametric panels show the other four best fits (each probing 200 random models, see Sec.\ref{sec:SEDmodeling}). The dashed and dotted lines show starforming and quiescent threshold, respectively, proposed by \cite{Pacifici16}.}
\label{fig:SIMBAgal}
\end{figure}

We calculate the SED of the simulated galaxy and convolve the redshifted, k-corrected SED with transmission curves of filters that can be found in CEERS pointing 1 (all HST, all NIRCam and F770W, F1000W, F1280W, F1500W, F1800W and F2100W MIRI).
As an error of the flux, if possible, we assumed values of errors of the most massive, MIRI detected QGC in observed sample. 
Otherwise, we used 10\% of the flux as an error.
We use this set of fluxes as input for the exact same CIGALE run as the \textit{MIRI run} described in Sec.~\ref{sec:SEDmodeling} and in Tab.~\ref{TAB:CIGALE_inp}. 

We compare the SED fit from CIGALE between \textit{MIRI} and \textit{no-MIRI runs} (see top panel in Fig.~\ref{fig:SIMBAgal}).
Without MIRI points, CIGALE estimates the A$_V$ by $\sim$1 mag higher than when MIRI points are included. 
Furthermore, age estimation from 3305, 3138 and 3062 Myr with MIRI points drops to 1713, 501, and 1737 Myr when MIRI is not used, for DelayedBQ, NonParametric and Regulator runs respectively. 

Now, we focus only on \textit{MIRI run} only.
We report a fairly consistent M$_\star$ estimate. 
In all three runs the estimated mass was lower than the real one in SIMBA, by a factor of $\sim$2 (0.2 dex) for DelayedBQ and by a factor of $\sim$3 (0.4 dex) for NonParametric and Regulator. 
The estimated SFR in each run is around $\sim$10~M$_\odot$/yr, which is $\sim$2-3 order of magnitude higher than the simulated value. 
While this is a large deviation, in each run the galaxy is considered quenched by every criterion, even the most conservative one. 
Finally, in SIMBA simulation, our galaxy formed at $z>7$.
However, in CIGALE runs the estimated formation redshift is $\sim$6, 4.5 and 4 for DelayedBQ, NonParametric and Regulator, respectively.  

We then follow the method described in App.~\ref{APP:sSFHs} to prepare specific SFHs (sSFHs) for each run (see Fig.\ref{fig:SIMBAgal}).
We estimate the quenching moment $\tau_q$ and the quenching time T$_q$ the same way as in the main text. 
The simulated galaxy has quenched at $\tau_q\sim3400$ Myr, and the quenching lasted $\textrm{T}_q\sim500$ Myr.
For CIGALE results, only the Regulator SFH has a resolution sufficient to estimate these parameters. 
We use the five best fits (see Sec.~\ref{sec:SEDmodeling}) to calculate the standard deviation and use it as uncertainty.
We report $\tau_q = 3880 \pm325$ Myr and T${_q=425\pm 285}$~Myr.
Thus, we find a similar trend to the observed galaxies.
We can recover $\tau_q$, which is directly related to the $\Delta$T.
It converges to the same value with low scatter, and we can use it as an evolutionary tracer.
We can also recover the true value of T$_q$, but the scatter is large ($\sim67\%$).
We should not use T$_q$ based on only photometry studies.

\section{QGs control sample}\label{APP:UVJ}
As a sanity check, we compare our results to massive QGs found by \cite{Carnall23b} and nine QGs from \cite{Long24}.
In \cite{Carnall23b} the authors found 15 sources using HST/NIRCam observations; however, the positions of only three of them were observed by MIRI (IDs from Tab.~2 in \citealt{Carnall23}: ID8888 with 10$\mu m$, 12.8$\mu m$, 15$\mu m$ and 18$\mu m$ detections, ID29497 with 15$\mu m$ detection, and ID17318 with 15$\mu m$ upper limit).
In \cite{Long24} the authors found 44 QGCs using the empirical NIRCam colour-colour diagram. 
The position of only nine of them were observed by MIRI producing in total: 3 detections and 1 upper limit in 5.5$\mu m$, 4 detections and 1 upper limit in 7.7$\mu m$,
2 detections in 10, 12.8 and 18$\mu m$ and 1 detection and one upper limit in 15$\mu m$.

Firstly, the quiescent nature of all three sources from \cite{Carnall23b} agrees within all three runs of the SFH model and in both \textit{MIRI} and \textit{no-MIRI runs} according to the \cite{Koprowski24} MS criterion. 
We do not recover only two to three QGs (depending on the SFH model) from \cite{Long24}.
In total, we find 75-83\% of the QGs in these two works.

Using the sSFR quiescence criterion, only DelayedBQ classifies all three \cite{Carnall23b} objects as QGs, while Regulator and NonParametric models find object 8888 not quiescent.
Using the \textit{MIRI run} we recover four to six out of nine QGs defined in the \cite{Long24} sample, while with the \textit{no-MIRI run} this number drops to two to four. 
Thus, we find 50-75\% of the QGs found in other works.

We report the M$_\star$ are in perfect agreement with previous studies, up to $<0.3$ dex within all CIGALE runs. 
Finally, the inclusion of MIRI points strongly influences the A$_V$ for two \cite{Carnall23b} galaxies; A$_V$ decreases by up to$\sim$0.80 mag for galaxy ID29497 and by up to $\sim$0.4 mag for galaxy ID8888, depending on the SFH model.
For galaxy ID17318 (with one upper limit), the inclusion of MIRI does not change the A$_V$ estimation significantly (the change is $\pm0.1$ mag, depending on the SFH model).

\section{Bin ranges}
The bins and sub-bins ranges of M$\star$ and $\Delta$T are presented in Tab.~\ref{TAB:deltaTbins}. 
\begin{table*}[t] 
\centering
\caption{Summary of M$_\star$ and $\Delta$T ranges of bins used for Fig.\ref{fig:Av_T}.}
    \begin{tabular}{c c c c c}
    \hline
    \hline
    $\log$M$_\star$ range & $\Delta$T$_1$ & $\Delta$T$_2$ & $\Delta$T$_3$& $\Delta$T$_4$\\
    \hline
    \multicolumn{5}{c}{DelayedBQ \textit{no-MIRI run}}\\ 
    \hline
    \phantom{0}8.1 -- \phantom{0}9.8 & \phantom{0}73 -- \phantom{0}450 & \phantom{0}499 -- \phantom{0}826 & 
    \phantom{0}857 -- 1221 & 4141 -- \phantom{0}8083 \\
    \phantom{0}9.8 -- 10.7           & \phantom{0}19 -- \phantom{0}269 & \phantom{0}316 -- \phantom{0}488 & 
    \phantom{0}490 -- 1400 & 1813 -- \phantom{0}3561 \\
    10.7 -- 11.4                     & \phantom{0}20 -- \phantom{0}389 & \phantom{0}406 -- \phantom{0}538 & 
    \phantom{0}555 -- \phantom{0}829 & 1311 -- \phantom{0}4544 \\
    \hline  
    \multicolumn{5}{c}{DelayedBQ \textit{MIRI run}}\\ 
    \hline
    \phantom{0}8.1 -- \phantom{0}9.8 &\phantom{0}11 -- \phantom{0}358 &\phantom{0}503 -- \phantom{0}700 &
    \phantom{0}778 -- 1210 &1227 -- \phantom{0}9759 \\
    \phantom{0}9.8 -- 10.6           &\phantom{00}7 -- \phantom{0}163 &\phantom{0}354 -- \phantom{0}509 &
    \phantom{0}614 -- 1849 &2257 -- \phantom{0}6350 \\
    10.6 -- 11.7                     &\phantom{0}46 -- \phantom{0}481 &\phantom{0}509 -- \phantom{0}674 &
    \phantom{0}816 -- 2218 &2222 -- \phantom{0}4682 \\
    \hline  
    \multicolumn{5}{c}{NonParametric \textit{no-MIRI run}}\\  
    \hline
    \phantom{0}8.0 -- \phantom{0}9.4 &\phantom{0}19 -- \phantom{0}768 &1038 -- 3386 &3492 -- 7968 &
    8335 -- \phantom{0}9049 \\
    \phantom{0}9.4 -- 10.6           &\phantom{0}19 -- \phantom{0}288 &\phantom{0}320 -- 1283 &
    1737 -- 2234 &2438 -- \phantom{0}6494 \\
    10.6 -- 11.4                     &\phantom{0}23 -- \phantom{0}522 & \phantom{0}639 -- 1116 & 
    1585 -- 3042 & 4057 -- \phantom{0}5866 \\
    \hline  
    \multicolumn{5}{c}{NonParametric \textit{MIRI run}}\\  
    \hline
    \phantom{0}8.0 -- \phantom{0}9.2 &\phantom{0}23 -- \phantom{0}128 &\phantom{0}182 -- \phantom{0}857 &
    \phantom{0}948 -- 3504 &3876 -- 10746 \\
    \phantom{0}9.2 -- 10.4           &\phantom{0}27 -- \phantom{0}164 &\phantom{0}270 -- 1202 &
    1246 -- 2235 &2914 -- \phantom{0}6946 \\
    10.4 -- 11.6                     & \phantom{0}20 -- 1169          &1342 -- 2329 &
    2375 -- 4273 &4336 -- \phantom{0}7541 \\
    \hline  
    \multicolumn{5}{c}{Regulator \textit{no-MIRI run}}\\ 
    \hline
    \phantom{0}8.0 -- \phantom{0}9.9 &182 -- 1000                     &1066 -- 1548 &1609 -- 8013 &8547 -- \phantom{0}9037 \\
    \phantom{0}9.9 -- 10.7           &\phantom{00}8 -- \phantom{0}405 &\phantom{0}410 -- \phantom{0}946 &1076 -- 3198 &3836 -- \phantom{0}6280 \\
    10.7 -- 11.4                     &162 -- \phantom{0}287           &\phantom{0}328 -- \phantom{0}762 &816 -- 2994 &3394 -- \phantom{0}5655 \\
    \hline  
    \multicolumn{5}{c}{Regulator \textit{MIRI run}}\\ 
    \hline
    \phantom{0}8.0 -- \phantom{0}9.9 &\phantom{0}70 -- \phantom{0}454 &\phantom{0}493 -- 1123 &1197 -- 3678 &7536 -- 10509 \\
    \phantom{0}9.9 -- 10.7           &186 -- \phantom{0}524           &\phantom{0}684 -- 1839 &2251 -- 3101 &4353 -- \phantom{0}7646 \\
    10.7 -- 11.6                     &\phantom{0}73 -- \phantom{0}645 &\phantom{0}702 -- 1966 &2104 -- 3870 &4688 -- \phantom{0}5797 \\
    \hline
\end{tabular}
\tablefoot{Table is split into sections related to each of the CIGALE runs. The first column shows the range of M$\star$ of each bin. The rest of the columns present the $\Delta$T ranges in Myr for each of the sub-bins.}
\label{TAB:deltaTbins}
\end{table*}

\end{appendix}
\end{document}